\documentclass[manuscript,screen]{acmart}
\usepackage{multirow}
\usepackage{booktabs}
\usepackage{graphicx}
\usepackage{longtable}
\usepackage{svg}
\usepackage{xcolor}
\definecolor{light-gray}{gray}{0.95}

\AtBeginDocument{%
  }

\setcopyright{acmlicensed}
\copyrightyear{2025}
\acmYear{2025}
\acmDOI{XXXXXXX.XXXXXXX}

\acmConference[Conference acronym 'XX]{Make sure to enter the correct
  conference title from your rights confirmation emai}{June 03--05,
  2018}{Woodstock, NY}
\acmISBN{978-1-4503-XXXX-X/18/06}

\begin{document}


\title{Blockchain Data Analytics: A Scoping Literature Review and Directions for Future Research}

\author{Marcel Bühlmann}
\email{marcel.buehlmann@unifr.ch}
\orcid{0009-0000-1886-207X}
\author{Hans-Georg Fill}
\email{hans-georg.fill@unifr.ch}
\orcid{0000-0001-5076-5341}
\author{Simon Curty}
\email{simon.curty@unifr.ch}
\orcid{0000-0002-2868-9001}
\affiliation{%
  \institution{University of Fribourg}
  \city{Fribourg}
  \country{Switzerland}
}

\renewcommand{\shortauthors}{Bühlmann, Fill and Curty}

\begin{abstract}
Blockchain technology has rapidly expanded beyond its original use in cryptocurrencies to a broad range of applications, creating vast amounts of immutable, decentralized data. As blockchain adoption grows, so does the need for advanced data analytics techniques to extract insights for business intelligence, fraud detection, financial analysis and many more. While previous research has examined specific aspects of blockchain data analytics, such as transaction patterns, illegal activity detection, and data management, there remains a lack of comprehensive reviews that explore the full scope of blockchain data analytics. This study addresses this gap through a scoping literature review, systematically mapping the existing research landscape, identifying key topics, and highlighting emerging trends. Using established methodologies for literature reviews, we analyze 466 publications, clustering them into six major research themes: illegal activity detection, data management, financial analysis, user analysis, community detection, and mining analysis. Our findings reveal a strong focus on detecting illicit activities and financial applications, while holistic business intelligence use cases remain underexplored. This review provides a structured overview of blockchain data analytics, identifying research gaps and proposing future directions to enhance the field's impact.
\end{abstract}




\maketitle

\section{Introduction}
\label{Introduction}

Blockchain technology is used to implement a multitude of use cases beside the initial cryptocurrency implementation of Bitcoin and is transforming industries as well as academia \cite{akcora_blockchain_2018, banafa_introduction_2023, nakamoto_bitcoin_2008, fill_blockchain_2019}. Platforms like Ethereum, that extend the decentralization, traceability and immutability characteristics with smart contract concepts, foster innovative solutions for supply chains, medical record storage, energy trading and many more~\cite{singh_review_2023, wang_blockchain-enabled_2019, petroni_big_2018, cunha_blockchain_2022, choubey_energytradingrank_2019}.


Simultaneously, data analytics has become crucial for enterprises, shifting decision-making from intuition to data-driven strategies to improve competitive performance and strategic innovation~\cite{haque_sazu_can_2022, liu_challenges_2018}. Data analytics itself has transformed in three phases~\cite{liu_challenges_2018,haque_sazu_can_2022}: The first phase focuses on historical data collection and reporting. The second, Business Intelligence, provides insights via dashboards but lacks prediction. The current third phase leverages Big Data, AI, and machine learning for predictive and prescriptive analytics.
The field of data analytics has evolved from basic data processing to a sophisticated tool that drives strategic decision-making, while unleashing this potential requires overcoming challenges related to skills, culture, and data management \cite{sedkaoui_how_2020, haque_sazu_can_2022, liu_challenges_2018}.


Due to the full record of all transactions in a blockchain, it offers enormous potential for data analytics~\cite{akcora_blockchain_2018, marsalek_tackling_2019}. Insights gained through applying analytics methods to blockchain data may be used to detect malicious activities, analyze customer behavior and user networks, or derive financial insights and business performance indicators such as revenue, price or cost indicators~\cite{dillenberger_blockchain_2019}.
Especially for companies integrating blockchains in their business models, it's necessary to retrieve data on the performance of their blockchain-based value activities in order to enable decision making and controlling \cite{curty_domain-specific_2023}. Further, applications that build upon the trust properties of blockchain, such as insurance applications, are in need for effective data analytics frameworks and tools to make informed data-based decisions~\cite{elsheikh_blockchain_2022}.


Over time, a variety of approaches to blockchain data analytics have evolved, addressing aspects of data collection and management, as well as predictive and prescriptive analytics~\cite{balaskas_analytical_2018,khan_graph_2022,singhal_coalescence_2021}. In addition, a multitude of literature surveys have been conducted in the context of blockchain data analytics, mostly covering specific topics such as the detection of illegal activities, e.g.~\cite{li_survey_2022,turner_analysis_2020,han_blockchain_2021}, analysis of transactions~\cite{pavithran_survey_2018,wu_analysis_2021,liu_knowledge_2021}, or the de-anonymization of identities on blockchains and digital forensics, e.g.~\cite{masud_review_2021,xi_review_2020,kebande_review_2022,srivasthav_study_2021}. Whereas surveys on blockchain data analytics in general exist, such as the one by~\citet{hou_survey_2021} in 2021 on general blockchain data analysis, by~\citet{wei_survey_2022} in 2022 on blockchain data management systems, or by \citet{huang_survey_2021} in 2021 on general theoretical and analytical models of blockchain, these either pre-suppose certain categories for analytics (e.g. security, privacy, performance and price prediction~\cite{hou_survey_2021}), do not focus on the goals and purpose of data analytics, e.g.~\cite{wei_survey_2022}, or shift the focus back to data analytics for detecting illegal activities and risks~\cite{huang_survey_2021}. What is missing so far is a survey on blockchain data analytics that takes a broad, exploratory perspective in order to determine the scope of this field. As shown by recent contributions in marketing research, blockchain data has the potential to give valuable insights for business use cases, which seems to be rather unexploited so far~\cite{hanneke_decoding_2024}. We were therefore interested in finding out which \emph{topics} in blockchain data analytics have so far been covered, how they have \emph{evolved over time}, and in which \emph{outlets} research on blockchain data analytics is published as well as the leading institutions operating in this field.

Following the typology of literature reviews by~\citet{pare_synthesizing_2015}, we therefore opted for a \emph{scoping review} of the literature on blockchain data analytics including assessing the quality of the investigated papers and identify gaps in the existing literature, as recommended by \citet{daudt_enhancing_2013}. 
In addition, we reverted to the guidelines proposed by \citet{webster_analyzing_2002} and \citet{kitchenham_guidelines_2007} for determining the steps in the search protocol as well as our own prior experiences in systematic literature reviews in the context of blockchains~\cite{curty_design_2023}.
In a first pre-study we found that only few outlets exist that focus on blockchain-specific topics and that to the best of our knowledge no outlets exist so far that focus specifically on blockchain data analytics. Therefore, we derived the first research question to be answered by the literature review as follows:\\
\indent \textit{RQ1: In which outlets is research on blockchain data analytics published and by which institutions?}\\
The answer to this research question will permit to identify the main venues where topics on blockchain data analytics are currently being discussed, as well as the institutions and countries of the involved authors, thus providing guidance for researchers aiming to publish new contributions.

Data analytics itself is not a means to an end. Rather, it is embedded in what is typically referred to as \emph{business intelligence}, which provides decision support for specific objectives in the context of business activities in a given domain to satisfy specific information needs~\cite{grossmann_fundamentals_2015,skyrius_business_2021}. The objectives of data analytics are thus of primary importance also in the context of blockchain data analytics. In contrast to previous literature reviews in blockchain data analytics, we again take an exploratory perspective here, without pre-supposing particular topics of analytics or objectives. Therefore we formulate the second research question as follows:\\
\indent \textit{RQ2: What are the major topics of blockchain data analytics?}\\
By answering this research question, we intend to describe the full scope of current topics in blockchain data analytics. This will permit to position contributions to blockchain data analytics research and lay the foundation for identifying missing topics such as intended by a scoping review~\cite{pare_synthesizing_2015}.

In relation to the second research question, our investigation will further examine the evolution of these topics through the passage of time. This will facilitate the identification of trends in blockchain data analytics since the genesis of Bitcoin. Therefore the third research question is formulated as:\\
\indent \textit{RQ3: How did the identified topics in blockchain data analytics evolve over time?}\\
By clustering all the papers of the review set in the topics and recording when they were published, it will be possible to discover potential trends in terms of the number of publications per year over time.

Data analytics should be part of a broader business intelligence strategy to support decision-making. Business intelligence has three key perspectives~\cite{grossmann_fundamentals_2015}: production (product offerings and operations), customer (behavior and product usage), and organizational (business processes). Since most blockchain data analytics research is narrow, this study examines unexplored topics within these perspectives.
Therefore we derive the fourth research question:\\
\indent \textit{RQ 4: Which topics in blockchain data analytics research are currently missing in comparison to the general scope of business intelligence?}\\
For answering this research question, we will compare the topics elaborated for existing approaches in blockchain data analytics to the scope of business intelligence. This will permit to identify areas of business intelligence and data analytics that have so far not been covered.

Finally, we are interested in getting first insights into how different blockchain data analytics approaches work. For this we will take a classical input-processing-output view as common for information systems~\cite{boell_conceptualizing_2012,laudon_management_2014}. It shall be investigated, which parts of blockchain data are used, how these are processed and which results are obtained. This leads us to the fifth research question:\\
\indent \textit{RQ5: What datasets, data processing approaches and results are used in blockchain data analytics?}\\
As we are performing only a scoping literature review and not a systematic literature review of all existing sources, the answers to this research question will not be given for all retrieved literature sources. Rather, based on the existing extensive literature reviews on specific blockchain data analytics topics, it is expected that only a subset of all sources will be used to illustrate the main approaches.

\textit{Our Contribution and Scope} \newline
The goal of this scoping literature analysis is to evaluate the current state-of-the-art in blockchain data analytics and provide an overview how data of a blockchain can be analyzed. We aim to capture the full breadth of blockchain data analytics in an exploratory style, without assuming any specific areas or topics. Furthermore, this review shall identify research gaps and future research directions in the field of blockchain data analytics, specifically in regard to analyzing blockchain data for enabling business use cases.


The remainder of this article is structured as follows: In Section \ref{Intorduction:Related-Work} we present related work and in Section \ref{Review}, we detail the methodology followed to conduct the systematic literature review. In Section \ref{Review:Descriptive_Analysis} we analyze the publications metadata and follow up by a content-based analysis of the topic clusters, where we determine the state of research for each topic cluster in Section \ref{Review:Topic_Categorization}. In Section \ref{Discussion} we discuss the findings by answering the research question and examine the future challenges along with future research directions. Finally, we conclude the paper in Section \ref{Conclusion}.

\section{Related Work on Blockchain Data Analytics}
\label{Intorduction:Related-Work}
Data analytics examines raw data to extract insights, combining statistics, computer science, and domain expertise. Key processes include data collection, preprocessing, modeling, and interpretation. As a multidisciplinary field, it integrates statistical methods, computational strategies, and AI to enhance decision-making \cite{moreira_general_2019}.
Data analytics can be broadly divided into four main categories. \textit{Descriptive analytics} is concerned with the identification, elucidation and summarization of a phenomenon. In contrast, \textit{diagnostic analytics} seeks to ascertain the underlying causes of a phenomenon. This entails exploratory data analysis of existing datasets or supplementary data collection, as required. \textit{Predictive analytics} extends beyond the comprehension of past and present phenomena to encompass the anticipation of future outcomes and the examination of potential scenarios by employing statistical and data mining methodologies. Finally, \textit{prescriptive analytics} advances to recommending the optimal course of action to achieve the desired future outcome by employing optimization and simulation to enhance business processes and achieve objectives in an effective manner \cite{moreira_general_2019, banerjee_data_2013}.
Big data analytics handles large datasets with volume, variety, and velocity, while data science provides advanced modeling tools. Applied in fields like healthcare and finance, techniques include statistical summarization, visualization, and machine learning \cite{moreira_general_2019}.
Data analytics now extends to blockchain, which produces vast, complex datasets. Specialized techniques are needed to extract insights, detect patterns, and support decision-making. Due to blockchain’s decentralized and immutable nature, analytics must adapt to challenges like transparency, security, and scalability \cite{huang_survey_2021}.

Within the field of data analytics of blockchain, multiple literature reviews and surveys have been performed -- see Table~\ref{tab:surveys}. Details on the search protocol and the identification of these surveys will be presented in Section~\ref{Review}.

\begin{table*}[!b]
    \footnotesize
  \begin{tabular}{p{0.03\textwidth}p{0.7\textwidth}    p{0.2\textwidth}}
    \toprule
    Year & Title & Focus\\
    \toprule
     2018 & Analytical Tools for Blockchain: Review, Taxonomy and Open Challenges \cite{balaskas_analytical_2018} & Tools \\ \hline
     2018 & A Survey on Analyzing Bitcoin Transactions \cite{pavithran_survey_2018} & Transaction Analysis \\\hline
     2019 & A Survey of Anonymity of Cryptocurrencies \cite{amarasinghe_survey_2019} & Anonymity \\\hline
     2020 & A Review on Data Analysis of Bitcoin Transaction Entity \cite{xi_review_2020} & De-Anonymisation \\\hline
     2020 & Analysis Techniques for Illicit Bitcoin Transactions \cite{turner_analysis_2020} & Illegal Activity Detection \\\hline
     2021 & A Review of Digital Forensics Framework for Blockchain in Cryptocurrency Technology \cite{masud_review_2021} & Digital Forensics \\\hline
     2021 & Analysis of cryptocurrency transactions from a network perspective: An overview \cite{wu_analysis_2021} & Transaction Analysis \\\hline
     2021 & A Survey of State-of-The-Art on Blockchains \cite{huang_survey_2021} & General \& Data Management \\\hline 
     2021 & A survey on blockchain data analysis \cite{hou_survey_2021} & General Analysis Approaches \\\hline
     2021 & A survey on public blockchain-based networks: structural differences and address clustering methods \cite{shin_survey_2021} & Address Clustering \\\hline
      2021 & Blockchain Abnormal Transaction Behavior Analysis: a Survey \cite{han_blockchain_2021} & Illegal Activity Detection \\\hline
    2021 & Coalescence of Artificial Intelligence with Blockchain: A Survey on Analytics Over Blockchain Data in Different Sectors \cite{singhal_coalescence_2021} & Artificial Intelligence \\\hline
    2021 & Knowledge Discovery in Cryptocurrency Transactions: A Survey \cite{liu_knowledge_2021} & Transaction Analysis \\\hline
    2021 & Revealing and Concealing Bitcoin Identities: A Survey of Techniques \cite{bergman_revealing_2021} & De-Anonymisation \\\hline
    2021 & Visualization of Blockchain Data: A Systematic Review \cite{tovanich_visualization_2021} & Visualization \\\hline
     2021 & Study of Blockchain Forensics and Analytics tools \cite{srivasthav_study_2021} & Digital Forensics \\\hline
    2022 & A Survey of Blockchain Data Management Systems \cite{wei_survey_2022} & Data Management \\\hline     
    2022 & A Survey on Ethereum Illicit Detection \cite{li_survey_2022} & Illegal Activity Detection \\\hline
    2022 & A Systematic Review of Detecting Illicit Bitcoin Transactions \cite{lin_systematic_2022} & Illegal Activity Detection \\\hline    
    2022 & Graph Analysis of the Ethereum Blockchain Data: A Survey of Datasets, Methods, and Future Work \cite{khan_graph_2022} & Graph Analysis \\\hline
    2022 & Review of Blockchain Forensics Challenges \cite{kebande_review_2022} & Digital Forensics \\\hline
    2023 & A Survey on the Efficiency, Reliability, and Security of Data Query in Blockchain Systems \cite{zhang_survey_2023} & Data Query    
    \\
    \bottomrule
  \end{tabular}
\caption{Previous reviews and surveys regarding aspects of blockchain data analytics ordered by year.}
  \label{tab:surveys}
\end{table*}


Blockchain data analytics is regarded as a sub-field of blockchain research. However, in a survey by \citet{huang_survey_2021} in 2021 only one previous review in Chinese on the status, trends, and challenges of blockchain data analytics was found, which underlines early stage of the field and the need for further research. 


According to \citet{moreira_general_2019}, the first step to any kind of data analysis is data collection and extraction followed by the analysis itself. Therefore, we first investigated the survey of \citet{wei_survey_2022} that explores data storage, management, and extraction, with the key-components for optimization, in different blockchains.
Subsequently, \citet{zhang_survey_2023} examined the different approaches of blockchain data querying and analyze blockchain data queries from three perspectives: efficiency, verification, and privacy. The study contributes by identifying ongoing challenges such as storage efficiency and query optimization.

Several previous surveys discuss the analysis of blockchain transactions, respectively the transaction or user graph, like 
\citet{hou_survey_2021}'s survey on blockchain data analytics. They cluster approaches into security, privacy, performance, and price prediction analysis and highlight blockchain data analytics, especially data mining, as a growing field but note that their own review is limited to only 50 studies from 2016–2020.
\citet{amarasinghe_survey_2019} explored anonymity in cryptocurrencies, showing that Bitcoin does not fully satisfy all anonymity properties. They emphasize the need for additional privacy measures like mixing protocols and cryptographic solutions. The authors call for further research, noting that anonymity methods are underdeveloped.
\citet{pavithran_survey_2018} review Bitcoin transaction analysis, categorizing studies into anonymity, wealth accumulation, user behavior, network, and price trends. They emphasize the need for more research on de-anonymization due to growing transaction data. Their categorization of analysis approaches is limited to Bitcoin, which heavily influences their derived analysis topics since no platforms capable of smart contracts are included.
\citet{xi_review_2020} review methods for identifying transaction entities using the Bitcoin transaction graph, focusing on heuristic algorithms, descriptive statistics, network, and visual system analysis.
\citet{bergman_revealing_2021} review how Bitcoin identities can be revealed and concealed. They discuss multiple identity-revealing methods and the respective concealing techniques. However, most can be countered by address clustering. The authors conclude that identity concealment and revealing are in constant competition.
\citet{shin_survey_2021} compare Bitcoin and Ethereum regarding accounts, transactions, and double-spending mitigation techniques. Address clustering in Bitcoin relies on addresses, transactions, or entire networks, while in Ethereum clusters are also created for smart contracts and their users. However, the survey lacks a clear methodology for filtering relevant literature, making its completeness unverifiable.

In the survey of \citet{han_blockchain_2021}, the authors explore detecting abnormal transaction behavior from two perspectives: smart contract analysis and broader transaction behavior. They identify methods like feature-based smart contract analysis and behavior classification. Unfortunately, the survey lacks a clear methodology and focuses only on a small subset of data analysis approaches.
\citet{lin_systematic_2022} review methods for detecting illicit Bitcoin transactions. They identify three main approaches: supervised learning, unsupervised learning, and topology-based methods. The authors highlight graph-based Bitcoin flow analysis as a promising area for future research.
\citet{li_survey_2022} studies illicit transaction detection on Ethereum with general strategies like supervised and unsupervised learning on transaction networks and specialized strategies that use machine learning and data mining based on code and account features.
\citet{turner_analysis_2020} review illicit Bitcoin transaction analysis, covering the legal landscape, analytical techniques, and ransomware countermeasures. They discuss cryptocurrency regulations, financial intelligence units, and various analysis methods. The authors conclude that individual techniques offer limited benefits but, when combined, significantly enhance law enforcement efforts.
\citet{wu_analysis_2021} examine transaction graph analysis methods for cryptocurrency, covering network properties, market effects, and detection techniques for entities, patterns, and illicit activities. They acknowledge extensive existing research but highlight gaps in dynamic network analysis and online learning.
\citet{masud_review_2021} explore digital forensics in cryptocurrencies, outlining a framework based on existing forensic models. The authors note that existing basic forensic methods lack solutions for evidence preservation in legal cases and universal applicability across cryptocurrencies.
\citet{kebande_review_2022} review nine blockchain forensic tools, covering transaction analysis, custody tracking, and log analysis, but note the lack of procedures for blockchain forensics. \citet{srivasthav_study_2021} examine 13 tools, distinguishing between investigation depth, risk detection, real-time monitoring, and audit capabilities, highlighting the limitations of open-source tools.
\citet{balaskas_analytical_2018} propose a taxonomy for blockchain data analytics tools. They apply this framework to existing tools, noting overlaps and blockchain-specific limitations. However, the taxonomy is very specific, focusing mainly on transactions and participants and overlooking other data.

\citet{liu_knowledge_2021} analyze blockchain transaction data, focusing on tracking transactions across entities and tracking countermeasures. They explore complex network analysis to study transaction patterns and individual behavior using machine learning for anomaly detection. The authors also review visualization tools, highlighting how blockchain transparency offers insights into human socio-economic behavior.
\citet{tovanich_visualization_2021} survey blockchain visualization tools as well. They identify key uses, including transaction analysis, cybercrime detection, and market insights. Their study finds that visualization mainly aids in communicating analysis results rather than directly supporting the analysis of blockchain data.

The survey conducted by \citet{khan_graph_2022} discusses Ethereum transaction graph analysis, and how comparable the conducted research is, covering multiple approaches for tasks like node classification, anomaly detection, and price prediction. The analysis of decentralized apps, decentralized finance, and multi-layer networks is identified as future research directions for blockchain data analytics.
\citet{singhal_coalescence_2021} explore AI-driven blockchain data analytics across sectors distinguishing between on-chain and transactional data analysis. They highlight the lack of research on AI-blockchain integration and its vast potential but note their survey lacks a holistic, systematic approach.

\section{Applied Search Protocol and Review Process}
\label{Review}
For determining the current state-of-the-art we conduct a scoping review according to \citet{pare_synthesizing_2015}, since we focus on the \emph{breadth} of available literature rather than providing an in-depth analysis of all data analysis techniques. For this we employ the following multi-step research methodology adhering to the guidelines by \citet{webster_analyzing_2002}, \citet{kitchenham_guidelines_2007} and \citet{vom_brocke_standing_2015} for structured literature reviews.
\citet{webster_analyzing_2002} assume that the relevant literature of a research topic is present in the leading journals. Only after the relevant literature has been identified in the leading publication outlets, the citations and references should be retrieved. However, for blockchain data analytics no central set of leading publication outlets could be identified, which renders the proposed starting point impractical.
In contrast, \citet{kitchenham_guidelines_2007} recommend a preliminary literature search to identify existing reviews and refine keywords. They suggest using database and manual searches across journals and conference proceedings and emphasize removing duplicates to prevent bias in the analysis.
\citet{vom_brocke_standing_2015} outline the literature review process in three phases: search, selection, and analysis, which can occur sequentially or iteratively. Finding literature in online databases is challenging due to the vast number of journals and inconsistent terminology, requiring searches across multiple databases. They recommend testing keywords and exploring publication outlets directly. Additionally, they propose a backward-forward search to ensure a comprehensive review beyond the outlets in the initial databases.

\begin{figure}[h]
  \centering
  \includegraphics[width=\linewidth]{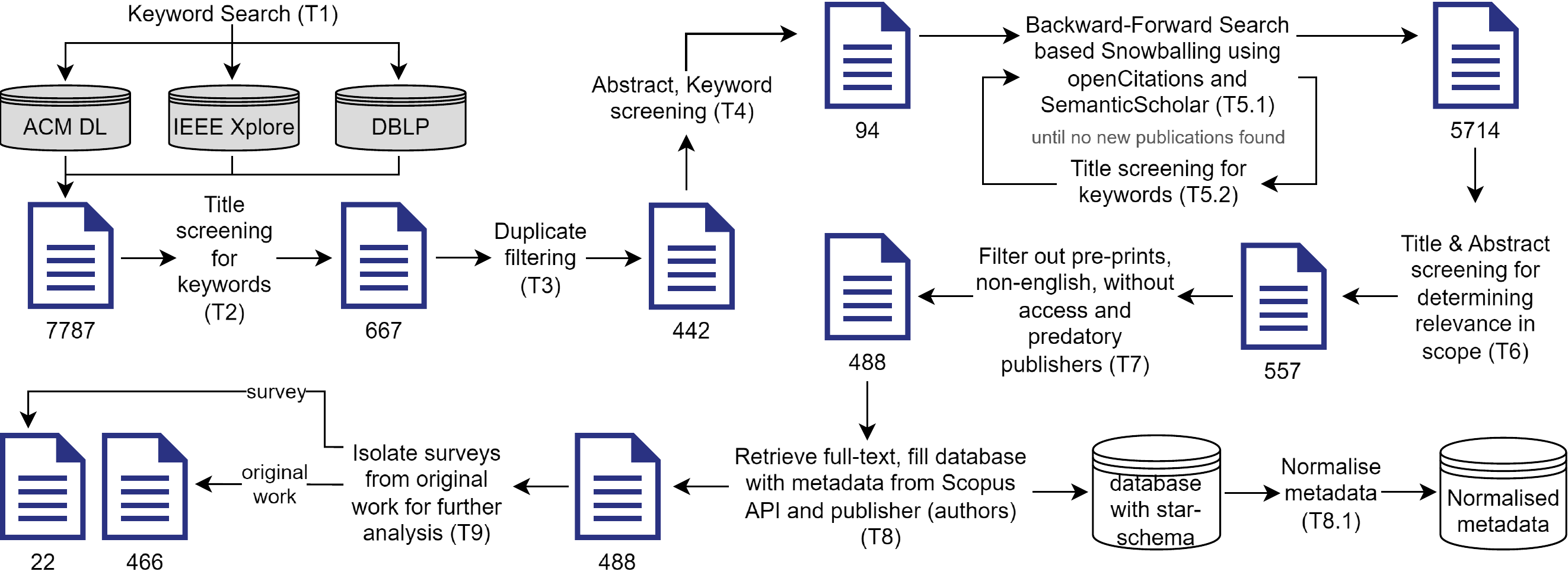}
  \caption{The diagram of the applied systematic literature search, retrieval and review process with individual activities and number of publications per step. The literature retrieval process is supplemented with the metadata-retrieval and metadata normalization in a relational database.}
  \Description{Process model of the data retrieval process}
  \label{figures:Data_retrieval_process}
\end{figure}

Building on the guidelines from \citet{webster_analyzing_2002}, \citet{kitchenham_guidelines_2007}, and \citet{vom_brocke_standing_2015}, we combined their approaches for our study. Our review starts with details on a database search, followed by a backward-forward search (BFS) to compile a comprehensive set of blockchain data analytics literature. The retrieved publications were analyzed and clustered into research directions using the literature coding process proposed by \citet{hruschka_reliability_2004}. Finally, we describe the topic clusters in detail. The process is illustrated in Figure \ref{figures:Data_retrieval_process}.

We first performed searches in the IEEE Xplore, ACM Digital Library and DBLP databases (T1). These databases were chosen as the contained publications exhibit minimal quality standards and the databases' scope matches the subject area of our literature review \cite{kitchenham_guidelines_2007}. The subsequent backward-forward search ensures that also publications not available in the selected databases are considered by our review. The present literature search was conducted within the time span from the inception of Bitcoin in 2008 (01.01.2008) until the cut-off date of on February 02, 2024.

In addition to the authors' domain-specific knowledge, a pre-study played a significant role in the selection of keywords \cite{vom_brocke_standing_2015,kitchenham_guidelines_2007}. 
To retrieve suiting research, we combined the keywords \textit{blockchain, smart contract, distributed ledger} with \textit{data analytics, analytics, business intelligence}. This led to the search strings \textit{Blockchain Analytics}, \textit{Blockchain Data Analytics}, \textit{Blockchain Business Intelligence}, \textit{Distributed Ledger Analytics} and \textit{Smart Contract Analytics}. The detailed search queries are listed in Table \ref{tab:Search-Queries-DB}. 

\begin{table*}[]
\footnotesize
\centering
\begin{tabular}{ p{0.13\textwidth}  p{0.83\textwidth} }
\toprule
\textbf{Database}         & \textbf{Search Query} \\
\midrule
\multirow{5}{*}{\parbox{0.12\textwidth}{IEEE Xplore (Command Search)}} & \texttt{("Full Text \& Metadata":Blockchain Analytics)} with filter applied \texttt{2008-2024}\\ \cline{2-2} 
                        & \texttt{("Full Text \& Metadata":Blockchain Data Analytics)} with filter applied \texttt{2008-2024}\\ \cline{2-2}
                        & \texttt{("Full Text \& Metadata":Blockchain Business Intelligence)} with filter applied \texttt{2008-2024}\\ \cline{2-2}
                        & \texttt{("Full Text \& Metadata":Distributed Ledger Analytics)} with filter applied \texttt{2008-2024}\\\cline{2-2}
                        & \texttt{("Full Text \& Metadata":Smart Contract Analytics)} with filter applied \texttt{2008-2024}\\ \hline
\multirow{5}{*}[-35pt]{\parbox{0.12\textwidth}{ACM Digital Library (Advanced Search Query Syntax)}} & \texttt{"query": { Title:(Blockchain AND "Analytics") OR Fulltext:(Blockchain AND "Analytics") }
"filter": { E-Publication Date: (01/01/2008 TO 02/07/2024), ACM Content: DL }}\\\cline{2-2}
                        & \texttt{"query": { Title:(Blockchain AND "Data Analytics") OR Fulltext:(Blockchain AND "Data Analytics") }
"filter": { E-Publication Date: (01/01/2008 TO 02/07/2024), ACM Content: DL }}\\ \cline{2-2}
                        & \texttt{"query": { Title:(Blockchain AND "Business Intelligence") OR Fulltext:(Blockchain AND "Business Intelligence") }
"filter": { E-Publication Date: (01/01/2008 TO 02/07/2024), ACM Content: DL }}\\\cline{2-2}
                        & \texttt{"query": { Title:("Distributed Ledger" AND Analytics) OR Fulltext:("Distributed Ledger" AND Analytics) }
"filter": { E-Publication Date: (01/01/2008 TO 02/07/2024), ACM Content: DL }}\\\cline{2-2}
                        & \texttt{"query": { Title:("Smart Contract" AND Analytics) OR Fulltext:("Smart Contract" AND Analytics) }
"filter": { E-Publication Date: (01/01/2008 TO 02/07/2024), ACM Content: DL }}\\ \hline
\multirow{5}{*}{\parbox{0.12\textwidth}{DBLP (URL Query)}}   & \texttt{DBLP.org/search?q=blockchain+analytics}\\\cline{2-2}
                        & \texttt{DBLP.org/search?q=blockchain+data analytics}\\\cline{2-2}
                        & \texttt{DBLP.org/search?q=blockchain+business intelligence}\\\cline{2-2}
                        & \texttt{DBLP.org/search?q=distributed ledger+analytics}\\\cline{2-2}
                        & \texttt{DBLP.org/search?q=smart contract+analytics}\\
                        \bottomrule
\end{tabular}
\caption{The executed search queries for each used database using the same keywords with database specific search query adaptations. The search queries correspond to the detailed execution queries executed in the advanced search options of the databases.}
\label{tab:Search-Queries-DB}
\end{table*}

Over all databases and search queries, 7787 results were retrieved. Those results were then assessed in activity T2 for relevance, primarily based on the title and keywords, and if necessary by reading the abstract.
We decided to focus on application-level aspects and thus filtered out publications that discuss network-level analysis, technical performance analysis, blockchain security analysis, or smart contract vulnerability analysis. 
In total, 667 results were deemed relevant, out of which 152 were identified as duplicates. The removal of the duplicates in activity T3 yielded 442 distinct results. 
Afterwards, all abstracts were read and each paper again assessed for relevance in activity T4. In this assessment the authors looked for original approaches for conducting data analytics with blockchain data. 
This second assessment reduced the initial set of 442 distinct publications to 94.

The previously assessed papers formed the starting set for the recursive backward-forward search in activity T5.1. The citations and references were retrieved from opencitations\footnote{opencitations.net} and semantic schoolar\footnote{semanticscholar.org}. For the backward-forward search, a python script was used, that takes the initial set as an input, with the DOI as identifier, and retrieves the respective reference and citation metadata from the APIs.
Once all citations and references were extracted, all newly added publications were evaluated for relevance in activity T5.2 based on title and keywords. All relevant publications were also considered in the following execution of the backward-forward search script. This iterative process was repeated until no new publications were found. The process was repeated 48 times in total and resulted in a set of 5714 publications.

The dataset of 5714 publications is again evaluated in activity T6 for relevance by two authors based on the title and the abstract, if deemed necessary. The evaluation was conducted individually and if conflicting assessments were made, an open discussion was initiated to resolve the disagreement. This analysis reduced the set to 557 publications. The set was further revised in activity T7 by filtering out non-peer reviewed literature published e.g. on arXiv, journals with predatory publishing methods\footnote{In light of concerns regarding the rigour of the review process and the resulting exclusion from the journal rating of the VHB (Verband der Hochschullehrerinnen und Hochschullehrer für Betriebswirtschaft) and Web of Science, all papers published by MDPI Journals have been excluded \cite{brainard_fast-growing_2024, hansen_eliminating_2024}.} and literature without access, resulting in 488 publications for further analysis.

In activity T8, we retrieved metadata attributes and the full-texts of all 488 publications. The following attributes were extracted: \emph{title}, \emph{authors}, \emph{publication year}, \emph{outlet}, \emph{outlet type}, \emph{institution}, \emph{institution city} and \emph{institution country}. The metadata were retrieved using the SCOPUS-API \footnote{dev.elsevier.com/documentation/ScopusSearchAPI.wadl} for all attributes except author information, with the DOI serving as the unique identifier. The author names were obtained directly from the respective publisher. The metadata was then loaded into a relational PostgreSQL database and normalized by \emph{author names}, \emph{publication year}, \emph{outlet}, \emph{institution}, \emph{institution city}, \emph{institution country} in activity T8.1. This was done to enable a multi-dimensional analysis of the metadata using a star schema. From the set of 488 publications, 22 were literature reviews or surveys that we separated from the original work in activity T9. This left us with 466 papers for conducting a content-based analysis.

\section{Descriptive Analysis Results of the Systematic Literature Review Process}
\label{Review:Descriptive_Analysis}
Research about blockchain data analysis is conducted all over the world in 51 different countries. The bulk of research is conducted by research institutions in China with 164 publications. This is followed by the United States with 42 and India with 39 publications, the rest is scattered around the world. For the top 5 publishing countries in the world, we list the top 3 institutions in Table~\ref{tab:Top-Institutions}. The extracted metadata shows that the \textit{Sun Yat-Sen University} publishes almost 3-times more than the next most publishing institution in China. The data also shows that apart from the Sun Yat-Sen University, no definitive leading institution can be identified.

Research results are usually published in academic literature, with various outlets and outlet types available to a researcher. Our analysis shows that most of the research is first published in conference proceedings (283 publications) while journal publications come second with 164 papers. Only a small portion of the conducted research was published as workshop proceedings (8 publications) or in books (11 publications). As illustrated in Table \ref{tab:Top-Outlets}, the most relevant outlets for research in blockchain data analytics are the \textit{International Conference on Blockchain and Trustworthy Systems} and the journal \textit{IEEE Access} with 15 and 13 publications respectively. The top 10 outlets further show, that no outlet is significantly more relevant than another, hence the relatively low number of publications per outlet with a base of 466 publications.

The development of blockchain data analytics approaches wasn't directly considered at the invention of Bitcoin. We can derive from the data illustrated in Figure \ref{figures:paper_per_topic_per_year}, that the first paper was published 3 years after the birth of Bitcoin in 2011. However, the topic gained traction around one of the first Bitcoin price spikes in 2017. Since then the number of publications on blockchain data analysis steadily increased, peaking in 2022 and this trend is expected to continue.

\begin{table*}[]
\footnotesize
\centering
\begin{tabular}{ p{0.2\textwidth}  p{0.55\textwidth}  p{0.18\textwidth} }
\toprule
\textbf{Country (\# of publications)}       & \textbf{Institution}         & \textbf{\# of publications} \\
\midrule
\multirow{3}{*}{China (164)}   & Sun Yat-Sen University         & 24\\ \cline{2-3} 
                        & Zhejiang University of Technology    & 9\\ \cline{2-3} 
                        & University of Electronic Science and Technology of China      & 6\\ \hline
\multirow{3}{*}{United States (42)} & IBM Research         & 3\\ \cline{2-3} 
                        & Georgia Institute of Technology      & 2\\ \cline{2-3} 
                        & Purdue University    & 2\\ \hline
\multirow{3}{*}{India (39)}   & Indian Institute of Technology Kanpur         & 4\\ \cline{2-3} 
                        & Birla Institute of Technology and Science & 3\\ \cline{2-3}
                        & Narsee Monjee Institute of Management Studies     & 3\\\hline
\multirow{3}{*}{Italy (24)}   & Università di Pisa         & 8\\ \cline{2-3} 
                        & Università degli Studi di Cagliari    & 4\\ \cline{2-3} 
                        & Università degli Studi di Perugia      & 4\\ \hline
\multirow{3}{*}{Australia (18)}   & Macquarie University         & 4\\ \cline{2-3} 
                        & Commonwealth Scientific and Industrial Research Organisation    & 2\\ \cline{2-3}
                        & Deakin University     & 2\\
                        \bottomrule
\end{tabular}
\caption{The list of the top 3 institutions of the top 5 countries based on the number of publications sorted by the number of papers published since 2011 and alphabetically when matching number of publications.}
\label{tab:Top-Institutions}
\end{table*}

\begin{table*}[]
\centering
\footnotesize
\begin{tabular}{ p{0.59\textwidth}  p{0.2\textwidth} p{0.0935\textwidth} p{0.025\textwidth} }
\toprule
\textbf{Outlet}       & \textbf{Outlet Type}    & \textbf{Publisher}     & \textbf{\#} \\
\midrule
        International Conference on Blockchain and Trustworthy Systems & Conference Proceeding & Springer & 15\\ \hline
        IEEE Access & Journal & IEEE & 13\\ \hline
        IEEE International Conference on Blockchain and Cryptocurrency & Conference Proceeding & IEEE & 10\\ \hline
        ACM World Wide Web Conference & Conference Proceeding & ACM & 9\\ \hline
        International Conference on Financial Cryptography and Data Security & Conference Proceeding & Springer & 9\\ \hline
        IEEE International Conference on Blockchain & Conference Proceeding & IEEE & 7\\ \hline
        IEEE International Conference on Big Data & Conference Proceeding & IEEE & 6\\ \hline
        Forensic Science International: Digital Investigation & Journal & Elsevier & 5\\ \hline
        IEEE International Conference on Data Mining Workshops & Conference Proceeding & IEEE & 5\\ \hline
        IEEE International Symposium on Circuits and Systems & Conference Proceeding & IEEE & 5\\

\bottomrule
\end{tabular}
\caption{The list of the top 10 publication outlets based on the number of publications with their respective publishers and number of publications since 2011.}
\label{tab:Top-Outlets}
\end{table*}


\section{Content Based Analysis - Major Topics in Blockchain Data Analytics}
\label{Review:Topic_Categorization}


\begin{figure}[h]
  \centering
  \includegraphics[width=\linewidth]{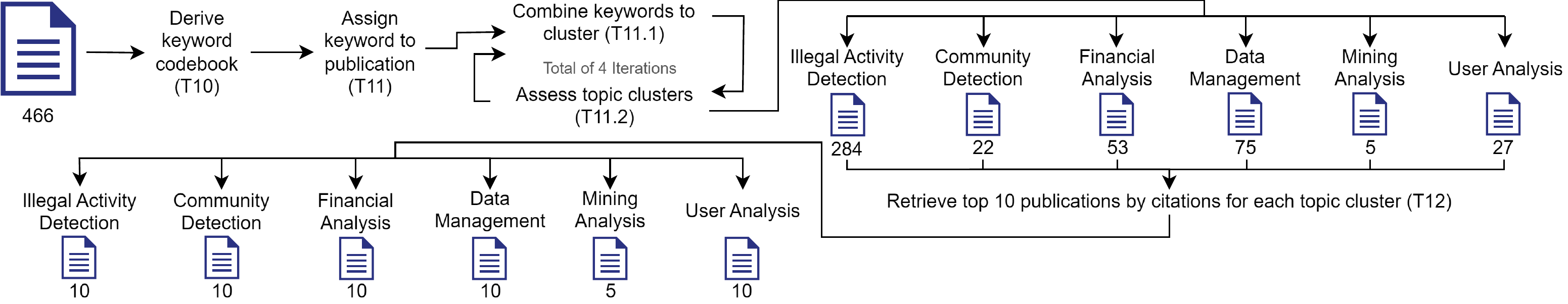}
  \caption{The diagram of the applied literature coding and iterative topic clustering process with the dataset size n=466.}
  \Description{Process model of the data retrieval process}
  \label{figures:clustering_process}
\end{figure}

For the analysis of the content of the selected publications we applied a clustering process as illustrated in Figure \ref{figures:clustering_process}. All publications (n=466) were clustered into topics in 5 iterations based in their analysis purpose. Thereby we followed the literature coding process proposed by \citet{hruschka_reliability_2004}. Two authors individually coded a subset of 160 publications consisting of 80 papers with at least 10 citations, and the 80 newest papers, using the developed codebook\footnote{doi.org/10.5281/zenodo.14970873}. In this first iteration, we achieveed a relative inter-rater agreement of 91.07\%, surpassing the reliability threshold of 85\% suggested by \citet{miles_qualitative_2014}. The coded subset was discussed by the coders in order to improve the common understanding. For every conflict a consensus between the coders could be achieved. The resulting keyword assignment was discussed by two authors to improve robustness. Related assigned keywords were then combined in topic clusters. The keywords were combined if the authors unanimously judged the topics as closely related. This process was repeated until keywords or topic clusters could no longer be combined.

Employing this clustering process, we assigned all 466 publications to 35 initial keywords and combined them manually into 6 topic clusters. As an example, we combined the codebook keywords \emph{crime detection},  \emph{fraud detection},  \emph{phishing detection} into the topic cluster  \emph{illegal activity detection}. The authors judged these keywords as closely related within the subject area of illegal activities. Similarly, the keywords \emph{cryptocurrency analysis}, \emph{financial analysis}, \emph{decentralized finance analysis} were combined under \emph{financial analysis}.
The finally derived topic clusters after five iterations were: \emph{Illegal Activity Detection}, \emph{Data Management}, \emph{Financial Analysis}, \emph{Community Detection}, \emph{User Analysis} and \emph{Mining Analysis}. For every previously defined topic cluster, we identified the most relevant key literature by retrieving the 10 most cited publications using the SCOPUS-API in activity T11. Although research on the correlation between citations and quality of research isn't conclusive, \citet{thelwall_which_2023} found a strong correlation between the number of citations and quality of research across all fields. We further assume, that highly cited publications are accepted in the respective research communities.

\begin{figure}[h]
  \centering
  \includegraphics[width=\linewidth]{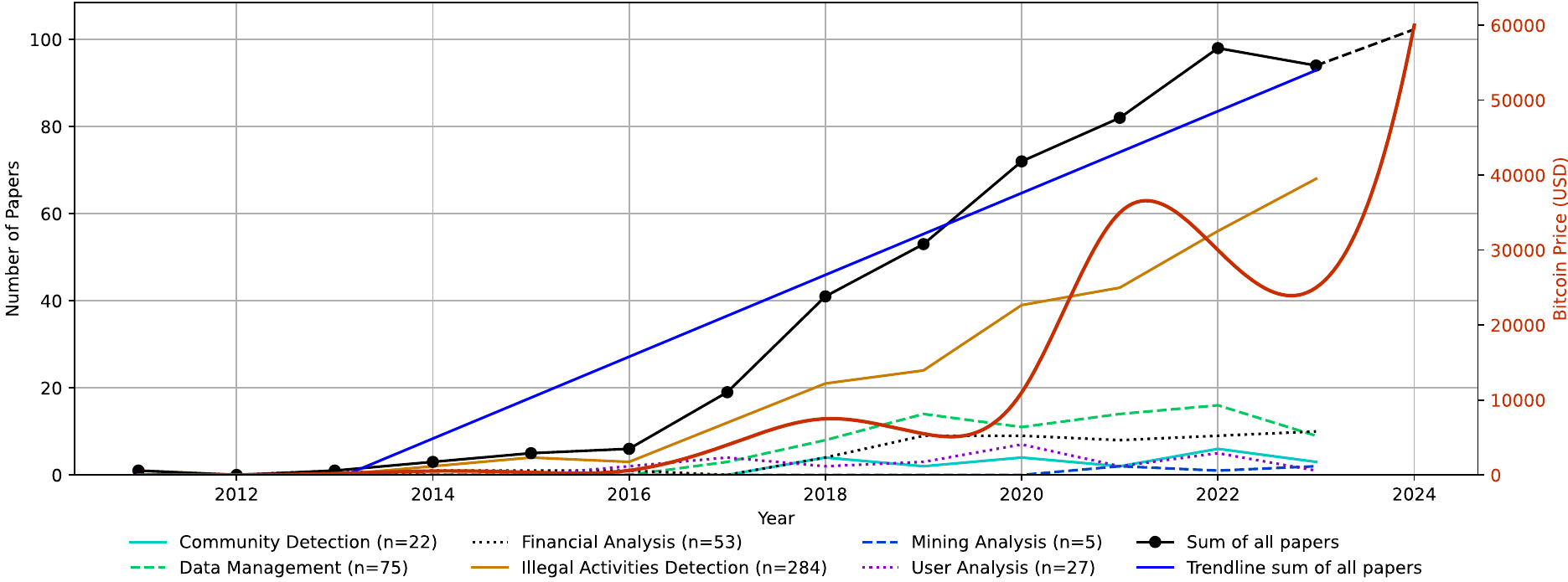}
  \caption{The evolution of the number of publications published per topic cluster derived during the literature review process and the total number of papers published per year between 2011 and the beginning of 2024 with n=466. The average exchange price (USD) of Bitcoin is added for referencing public awareness for blockchain, which might influence the number of papers published.}
  \Description{Number of papers per topic cluster published between 2011 and 2023 with n=466}
  \label{figures:paper_per_topic_per_year}
\end{figure}

In Figure \ref{figures:paper_per_topic_per_year} we illustrate the time-evolution of the topic clusters along with the average Bitcoin price (red line, right y-axis)\footnote{Average Bitcoin Price according to 99bitcoins.com/bitcoin/historical-price/ - accessed 04.12.2024 at 1730 CEST}. 
We observe that the interest of analyzing blockchain and cryptocurrency data emerged with a time lag after the inception of Bitcoin. Illegal activity detection was always the main focus of researchers based on the number of papers published. This gap opened exponentially from 2016 onward and reached its current peak in 2023. The interest in analyzing blockchain systems seems to be related to the price of Bitcoin. Therefore, we assume that in 2024 and 2025 the number of publications regarding blockchain data analytics will continue to increase. 

\subsection{Illegal Activity Analysis}
The unique anonymity features of blockchains make them highly attractive for criminals. Illegal activities are of various kinds and range from paying criminals with cryptocurrencies for an off-chain crime, money laundering to blockchain-based ponzi schemes.
This cluster includes topics like anomaly detection, blockchain forensics, de-anonymization, fraud detection, money laundering detection and transaction/ money tracking. The main focus of this topic cluster lies within the detection of illegal or malicious activities and to uncover the identities behind the malicious schemes. We list the identified and now discussed 10 key publications for illegal activity analysis in Table \ref{tab:Key-Paper-Illegal}.

\begin{table*}[]
\footnotesize
\centering
\begin{tabular} { p{0.05\textwidth}  p{0.25\textwidth} p{0.25\textwidth} p{0.25\textwidth} p{0.09\textwidth} }
\multicolumn{4}{l}{\textbf{Top 10 key publications for Illegal Activity Detection}}\\
\hline
\textbf{Ref (\#)}  & \textbf{Dataset} & \textbf{Data Processing} & \textbf{Result} & \textbf{Platform} \\ \hline
\cite{koshy_analysis_2014} (284) & Traffic of BTC network with associated Peer-IP's & Manual generalization and statistics & Mapped BTC addresses to (Peer-)IP's & BTC \\ \hline
\cite{jourdan_characterizing_2018} (210) & Entity labeled BTC dataset with transaction features & Spending heuristic and transitive closure operations & Classified transaction entities of BTC & BTC \\ \hline
\cite{alarab_competence_2020} (153) & Elliptic BTC labeled illicit transaction graph & Graph Convolutional Network & Prediction of illicit status for transactions & BTC \\ \hline
\cite{bartoletti_data_2018} (149) & Labeled ponzi and non-ponzi scheme BTC addresses & Random forest classification model & Detection of cryptocurrency-based frauds & BTC \\ \hline
\cite{biryukov_deanonymization_2019} (148) & Recorded network traffic of INV and ADDR messages & Transaction origin clustering bases on propagation times & Anonymity degree and attack security of a blockchain & BTC, ZC, D, M \\ \hline
\cite{farrugia_detection_2020} (112) & Set of labeled (illicit, normal) addresses & XGBoost classification model & Detection of illicit entities in unlabeled dataset, benchmark dataset & ETH \\ \hline
\cite{chan_ethereum_2017} (102) & ETH Transactions & Neo4j cypher queries & Stolen funds from the Gatecoin hack traced to addresses & ETH\\ \hline
\cite{chen_phishing_2020} (89) & Labeled phishing and non-phishing addresses & Graph Convolutional Network & Detection of ETH-based phishing scams & ETH \\ \hline
\cite{hu_transaction-based_2021} (79) & Transaction history smart contracts & Long sort-term memory network with feature extraction & Smart Contract type detection & ETH \\ \hline
\cite{mcginn_visualizing_2016} (72) & Densely connected multigraph transaction network & ForceAtlas2 visualisation algorithm of SigmaJS & Visual representation of the BTC transaction network & BTC \\ \hline
\end{tabular}
\caption{List of the top 10 key publications for the topic cluster Illegal Activity Detection included in the 55 key publications in Blockchain Data Analytics based on the number of citations (\#) per paper. Legend: BTC - Bitcoin, D - Dash, ETH - Ethereum, M - Monero, ZC - Zcash}
\label{tab:Key-Paper-Illegal}
\end{table*}

Several studies analyze anonymity risks in cryptocurrencies. \citet{koshy_analysis_2014} analyze Bitcoin anonymity by mapping addresses to IP addresses using P2P network traffic. Using their CoinSeer client, they collected five months of data and identified over 1,000 address-to-IP mappings through anomalous relay patterns. Their findings expose Bitcoin’s anonymity weaknesses and advocate for stronger privacy measures.
\citet{jourdan_characterizing_2018} examine Bitcoin entity characterization using a graph-based approach and machine learning to classify entities like exchanges and darknet markets. Their findings show that even weak attackers can infer entity types, while strong attackers achieve over 90\% accuracy, highlighting privacy risks and the need for better anonymity solutions.

\citet{alarab_competence_2020} introduce a novel approach that combines a Graph Convolutional Network with linear layers. In particular, they concatenate node embeddings with a hidden layer generated by a linear transformation of the node feature matrix, followed by a multi-layer perceptron. The experiment demonstrates an increased performance throughout all compared properties than previous graph neural network approaches. 
\citet{bartoletti_data_2018} use data mining and supervised learning to identify Bitcoin addresses linked to Ponzi schemes. They analyze real-world Ponzi transactions and frame detection as a binary classification problem, correctly classifying 31 out of 32 schemes with a 1\% false positive rate.

\citet{biryukov_deanonymization_2019} explore network-level privacy risks in Bitcoin, Dash, Monero, and Zcash, showing that an adversary can deanonymize transactions by analyzing propagation timing and network connections. They reveal links between addresses, exposing Bitcoin and Zcash transactions. The study highlights vulnerabilities even in privacy-focused cryptocurrencies, and suggests countermeasures like randomized message propagation.

\citet{farrugia_detection_2020} use machine learning to detect illicit Ethereum accounts, analyzing 2179 flagged and 2502 normal accounts. Their XGBoost model achieves 96.3\% accuracy and an AUC of 0.994. Key features include transaction time span, total Ether balance, and minimum received value. The study provides an effective detection method, identifies critical features, and offers a benchmark dataset for future research.

\citet{chan_ethereum_2017} analyze de-anonymization in Ethereum using graph analytics with Neo4j, focusing on transactions linked to hacks like Gatecoin and Coindash. They trace stolen funds and explore links to known entities, comparing Ethereum’s anonymity to Bitcoin’s. The study highlights Ethereum’s lack of unspent transaction outputs (UTXO) as a limitation and suggests combining blockchain graphs or web scraping for improved analysis.

\citet{chen_phishing_2020} detect phishing accounts using a Graph Convolutional Network with an autoencoder, modeling accounts and transactions as a graph and apply various embedding methods to labeled transaction data.
\citet{hu_transaction-based_2021} classify Ethereum smart contracts using transaction-based analysis. They analyzed 10,000 contracts, extracting 14 features and used an LSTM model for classification and anomaly detection. Their approach effectively identifies fraudulent contracts like Ponzi schemes, improving blockchain security.

\citet{mcginn_visualizing_2016} visualize the Bitcoin transaction graph. Their tool enables interactive exploration of transaction data and reveals attack patterns, including denial-of-service attacks, while statistical analysis highlights evolving attack vectors. The study concludes that visualization effectively uncovers illicit activities and aids expert analysis of Bitcoin transactions.

\subsection{Data Management}
Data management in blockchain systems focuses on efficiently storing, retrieving, analyzing, and securing data while addressing scalability and compliance challenges. According to \citet{moreira_general_2019}, this subject area therefore acts as an enabler for data analytics. The scalability of data storage and data retrieval performance is a key factor in order to establish blockchain as a viable technology for enterprises. This contributes to powerful blockchain data analytics, since the data bottleneck is mitigated. We list the identified 10 key publications for data management in Table \ref{tab:Key_Paper_Data}.

\begin{table*}[]
\footnotesize
\centering
\begin{tabular} { p{0.05\textwidth}  p{0.25\textwidth} p{0.25\textwidth} p{0.25\textwidth} p{0.09\textwidth} }
\multicolumn{4}{l}{\textbf{Top 10 key publications for Data Management}}\\
\hline
\textbf{Ref (\#)}  & \textbf{Dataset} & \textbf{Data Processing} & \textbf{Result} & \textbf{Platform} \\ \hline
\cite{paik_analysis_2019} (112) & Blockchain network data architectures & Qualitative assessment of data management approaches & Insights in data management of blockchain systems & BTC, ETH \\ \hline
\cite{li_etherql_2017} (110) & Full node synchronization & Query Layer & Generic query-layer for blockchain systems implemented on ETH & ETH \\ \hline
\cite{el-hindi_blockchaindb_2019} (105) & Full node synchronization & - & Blockchain-based database & ETH, HLS, HLF \\ \hline
\cite{bartoletti_analysis_2017} (103) & All OP\_RETURN transactions since BTC's origin block & Qualitative and quantitative assessment of OP\_RETURN content and transactions & Insights in OP\_RETURN usage & BTC \\ \hline
\cite{matzutt_quantitative_2018} (102) & BTC transactions holding additional data & Quantitative data examination and interpretation & Insight of available data in BTC & BTC \\ \hline
\cite{zheng_xblock-eth_2020} (101) & ETH blocks & Quantitative and qualitative assessment for dataset construction & Datasets for block and transaction, internal ether transaction, smart contracts, ERC20 and ERC721 Tokens & ETH \\ \hline
\cite{bandara_mystikoblockchain_2018} (71) & Mystiko blockchain & Miner, chain and storage services & Blockchain-based Big Data service & MY \\ \hline
\cite{vo_research_2018} (62) & Full node synchronization & - & Analysis how data analysis on blockchains can be supported and integrated in blockchains & - \\ \hline
\cite{peng_vql_2019} (46) & Full node synchronization & Query Layer & Verifiable Query Layer middleware for efficiency and data authenticity & ETH \\ \hline
\cite{amudha_dilated_2022} (42) & Full node synchronization & Dilated transaction access and retrieval method & Data retrieval service  & Unknown \\ \hline
\end{tabular}
\caption{List of the top 10 key publications for the topic cluster Data Management included in the 55 key publications in Blockchain Data Analytics based on the number of citations (\#) per paper. Legend: BTC - Bitcoin, ETH - Ethereum, HLF - Hyperledger Fabric, HLS - Hyperledger Sawtooth, MY - Mystiko}
\label{tab:Key_Paper_Data}
\end{table*}

\citet{paik_analysis_2019} discuss data management in blockchain systems, covering architecture, storage, analytics, and governance, comparing them to conventional databases.
\citet{li_etherql_2017} introduce EtherQL, a system that enhances Ethereum data querying by addressing storage and LevelDB limitations. It adds a query layer for advanced analysis, integrates with Ethereum clients, offers real-time synchronization, and provides APIs.
BlockchainDB is a database layer, proposed by \citet{el-hindi_blockchaindb_2019}, that leverages the blockchain's storage layer in order to enable classical data management techniques for blockchain systems. The authors implement their 3-tier architecture using Ethereum.

\citet{bartoletti_analysis_2017} examine Bitcoin’s OP\_RETURN opcode, which embeds metadata in transactions. Examining 1,887,708 OP\_RETURN transactions they identify 22 protocols. The study highlights spam attacks and scalability concerns, emphasizing OP\_RETURN's growing role in blockchain applications.
\citet{matzutt_quantitative_2018} analyze the impact of arbitrary content in Bitcoin on scalability, security, and legal compliance. They analyze its effects on storage, validation, and node operations, highlighting risks like illegal content and censorship challenges.
\citet{zheng_xblock-eth_2020} gather and process recent on-chain data from Ethereum and construct datasets. They label these refined datasets as XBlock-ETH, encompassing data on blockchain transactions, smart contracts, and cryptocurrencies. The newly constructed datasets are then analyzed by the authors in order to describe the state of Ethereum.

\citet{bandara_mystikoblockchain_2018} present Mystiko, a blockchain system optimized for big data storage and analytics. It enhances scalability and speed by leveraging Apache Cassandra, federated consensus, and sharding. Integrated with Elasticsearch, it enables efficient data retrieval. Performance tests show major gains in speed and efficiency.
\citet{vo_research_2018} discuss challenges in blockchain data management, including scalability, integration, security, and analytics in business networks. They advocate for database technologies, improved throughput, encryption, and AI-driven analytics. Overcoming these issues is crucial for broader enterprise adoption.
\citet{peng_vql_2019} show that blockchain queries are slow and lack authenticity verification. They propose a middleware that speeds up queries and ensures data integrity, directly confirming its efficiency and reliability.
\citet{amudha_dilated_2022} presents a dilated method for transaction retrieval using block identities and recursive structures. Controlled binary searches locate relevant blocks efficiently. Experiments with a NetSim emulator and IoT devices confirm its effectiveness.

\subsection{Financial Analysis}
The area of financial analysis of blockchains ranges from price predictions of cryptocurrencies or fees over token analysis, process mining and wealth analysis to visualization techniques for exchange services. This topic cluster is of particular interest to companies deploying applications on a blockchain, as financial variables have a direct impact on the profitability of a blockchain use case. We list the identified 10 key publications for financial analysis in Table \ref{tab:Key-Paper-Financial}.

\begin{table*}[]
\footnotesize
\centering
\begin{tabular} { p{0.05\textwidth}  p{0.25\textwidth} p{0.25\textwidth} p{0.25\textwidth} p{0.09\textwidth} }
\multicolumn{4}{l}{\textbf{Top 10 key publications for Financial Analysis}}\\
\hline
\textbf{Ref (\#)}  & \textbf{Dataset} & \textbf{Data Processing} & \textbf{Result} & \textbf{Platform} \\ \hline
\cite{li_analyzing_2020} (90) & Fundamental properties and features of blockchains & Qualitative feature and property analysis & Comparison between cryptocurrency properties & BTC, XRP, ETH, L \\ \hline
\cite{sin_bitcoin_2017} (80) & BTC time-series feature data & Artificial neural network with 5 Multi-Layered Perceptron & BTC price prediction model & BTC \\ \hline
\cite{yue_bitextract_2019} (68) & Transaction history processed in 1-to-1 transactions & Adaptive Network Standing Index and Business Proximity & Visual exchange network & BTC \\ \hline
\cite{jiang_cryptokitties_2021} (52) & ETH transactions & Transaction Network visualization and property analysis & Evolution of the CryptoKitties trading network and value & ETH \\ \hline
\cite{dustdar_elastic_2021} (48) & Smart contracts & Elasticity, performance analysis of execution behavior & Insights in behavior and performance of smart contracts & HFL \\ \hline
\cite{muhlberger_extracting_2019} (47) & Smart contract call transactions (Event Logs) with respective smart contract code & ProM Process Mining Workbench & Performance measures and sequence properties of blockchain-based processes & ETH \\ \hline
\cite{klinkmuller_mining_2019} (42) & ETH transactions and Event logs & ProM Process Mining Workbench & Directly-Follows graphs of a smart contract & ETH \\ \hline
\cite{motamed_quantitative_2019} (41) & Cryptocurrency transaction graphs and coin prices from CoinMarketCap & Graph analysis of constructed transaction graphs & CMTG, MTG measures cryptocurrency transaction graphs with correlation to the price & BTC, ETH, LC, D \\ \hline
\cite{pierro_influence_2019} (37) & Sub-set of block variables & Time-series analysis & Analysis of the Granger causality relationship & ETH \\ \hline
\cite{chen_traveling_2020} (32) & External/Internal transactions, smart contract information and calls & Token exchange graph analysis & Visualization, measures and properties of Token transaction graphs & ETH \\ \hline
\end{tabular}
\caption{List of the top 10 key publications for the topic cluster Financial Analysis included in the 55 key publications in Blockchain Data Analytics based on the number of citations (\#) per paper. Legend: BTC - Bitcoin, D - Dash, ETH - Ethereum, HLF - Hyperledger Fabric, L - Libra, LC - Litecoin, XRP - XRP}
\label{tab:Key-Paper-Financial}
\end{table*}

\citet{li_analyzing_2020} examine major cryptocurrencies, focusing on identity management, consensus mechanisms, and supply. Different cryptocurrencies offer varying features, balancing anonymity, security, decentralization, and stability while facing unique challenges and policy considerations. The study highlights their economic roles and calls for further research on market impact and regulation.
\citet{sin_bitcoin_2017} examine which Bitcoin features influence the next day price by employing a neural network approach. The applied neural network showed a great performance with price predictions for S\&P500 price predictions. The experiment conducted on real-world data demonstrated an accuracy of 58\% - 63\% with which an investor could make a profit of up to 85\%.

\citet{yue_bitextract_2019} introduce BitExTract, an interactive visual system for analyzing Bitcoin exchange transaction patterns. The system visualizes market evolution for exchanges and trading networks. BitExTract also compares transaction patterns across exchanges.
\citet{jiang_cryptokitties_2021} analyze the lifecycle of CryptoKitties, tracking five million transactions over three years. They identify four stages: primer, rise, fall, and serenity, noting that media attention drove growth, but oversupply, wealth concentration, and blockchain limitations led to decline. The study highlights challenges in game design and offers recommendations for creating sustainable blockchain games with long-term player engagement.

\citet{dustdar_elastic_2021} explore elastic smart contracts for adaptable IoT ecosystems like smart cities. Using virtual chains and glue contracts, elastic smart contracts adjust resources, quality, and costs dynamically. Tested on Hyperledger Fabric, they improve performance, efficiency, and scalability for smart infrastructure.

\citet{muhlberger_extracting_2019} create a framework to extract Ethereum smart contract data into event logs for process mining, enabling process monitoring and analysis. Their prototype addresses challenges like hexadecimal formats and timestamp approximations, showing how blockchain event logs enhance auditability and transparency. Future work will refine extraction and integrate off-chain data.
\citet{klinkmuller_mining_2019} propose a framework for mining blockchain processes to overcome the challenge of extracting data for process analysis. The framework includes a generator for logging code, an extractor for data in the XES format, and a manifest outlining logging requirements. They also introduce a technique for low-cost, high-throughput on-chain logging to condense data effectively.

\citet{motamed_quantitative_2019} examine transaction networks of five major cryptocurrencies, finding scale-free properties with dominant addresses. Transaction activity correlates with price movements, especially at market peaks. Ethereum and Z-Cash show higher node and edge repetition due to their architectures.
\citet{pierro_influence_2019} study how factors like pending transaction volume, the USD/Ether exchange rate, electricity prices, and miner activity affect Ethereum transaction fees. While some variables, like miner count, were correlated with gas price changes, no effect was found for others.
\citet{chen_traveling_2020} study the Ethereum ERC20 token ecosystem by analyzing transaction records and event logs to create three graphs: token creator, token holder, and token transfer. They propose an algorithm to detect relationships between tokens and accounts, identifying nearly 170,000 ERC20 tokens.

\subsection{User Analysis}
User analysis in blockchain focuses on examining interactions, behaviors, and roles within the network, using graph-based methods to uncover patterns. Advanced techniques like transaction subgraphs and graph neural networks enable better identity and behavioral analysis, overcoming challenges in large-scale networks. These studies highlight the differences between blockchain and traditional social networks, such as their smaller size, higher connectivity, and unique behaviors. Researchers aim to improve understanding of blockchain ecosystems, enhance system transparency with a focus of user behavior. We list the identified 10 key publications for user analysis in Table \ref{tab:Key-Paper-User}.

\begin{table*}[]
\footnotesize
\centering
\begin{tabular} { p{0.05\textwidth}  p{0.25\textwidth} p{0.25\textwidth} p{0.25\textwidth} p{0.09\textwidth} }
\multicolumn{4}{l}{\textbf{Top 10 key publications for User Analysis}}\\
\hline
\textbf{Ref (\#)}  & \textbf{Dataset} & \textbf{Data Processing} & \textbf{Result} & \textbf{Platform} \\ \hline
\cite{di_francesco_maesa_analysis_2017} (70) & BTC user (address) graph & Indegree outlier analysis with PS-Transaction detection & Infer anomalous user behavior in blockchain systems & BTC \\ \hline
\cite{sun_bitvis_2019} (55) & Clustered addresses with respective transactions & Relationship exploration based on circulation analysis & Constructed visual user relation graph & BTC \\ \hline
\cite{serena_cryptocurrencies_2022} (37) & Directed transaction network & Python Graph analysis library NetworkX & Graph analysis tool for statistical transaction graphs & BTC, DC, ETH, R \\ \hline
\cite{arroyo_dao-analyzer_2022} (25) & Real-time API connection to blockchains & - & Toolsuite for DAO analysis & ETH, GC, P, A \\ \hline
\cite{di_francesco_maesa_data-driven_2018} (20) & BTC transactions since the genesis block & Weighted directed hypergraph clustering algorithm & Clustered directed user graph & BTC \\ \hline
\cite{di_francesco_maesa_detecting_2017} (20) & BTC user graph based on transaction network & Graph analysis (Indegree frequency, diameter) & Transaction network properties and economical effect & BTC \\ \hline
\cite{bai_evolution_2022} (17) & Temporal transaction graph & Temporal UCG and CCG network analysis & Behavior analysis of addresses & ETH \\ \hline
\cite{shen_identity_2021} (17) & EOSG and ETHG dataset (EOS and ETH transaction graphs) & Graph neural networks and I\(^2\)BGNN model for graph classification & User behavior classification & EOS, ETH \\ \hline
\cite{lee_measurements_2020} (11) & ETH transaction network & Local and global network and graph analysis & Network features ETH and Token interaction graphs & ETH \\ \hline
\cite{mizerka_role_2020} (10) & MSCI Emerging Markets Index and BTC addresses with respective transactions & Regression analysis of BTC transactions & Insights on the role of BTC for investments & BTC \\
\hline
\end{tabular}
\caption{List of the top 10 key publications for the topic cluster User Analysis included in the 55 key publications in Blockchain Data Analytics based on the number of citations (\#) per paper. Legend: A - Arbitrum, BTC - Bitcoin, DC - Dogecoin, ETH - Ethereum, GC - Gnosis Chain, P - Polygon, R - Ripple}
\label{tab:Key-Paper-User}
\end{table*}

\citet{di_francesco_maesa_analysis_2017} examine anomalous Bitcoin transactions, focusing on pseudo-spam transactions that inflate address connectivity to conceal user behavior and user relationships. They identify transactions as spam, advertising, or de-anonymization attacks and suggest filtering them to improve Bitcoin's network structure, making it more small-world-like.
\citet{sun_bitvis_2019} present BitVis, an interactive tool for analyzing Bitcoin transactions and detecting suspicious activity using graph-based visualizations and filters. Built with Python, MySQL, and Neo4j, it enhances Bitcoin monitoring and forensic investigations.
\cite{serena_cryptocurrencies_2022} analyze cryptocurrency transactions and uncover interaction patterns in cryptocurrencies. They introduce a tool (DiLeNA) that visualizes transaction networks, revealing small-world properties across all cryptocurrencies offering insights into user interactions.
\cite{arroyo_dao-analyzer_2022} present a tool for analyzing activity and participation in Decentralized Autonomous Organizations. It tracks proposals, voting, and engagement trends using blockchain data. The goal is to support research on governance and sustainability.
\citet{di_francesco_maesa_data-driven_2018} analyze the Bitcoin user graph using a new clustering algorithm. They examine graph properties, confirming the \textit{rich get richer} phenomenon and the presence of central nodes.

\citet{di_francesco_maesa_detecting_2017} show that the network's structural properties stem from distinct and unusual patterns in the user graph. The conducted analysis suggests that the observed unusual patterns are likely caused by artificial user behaviors rather than typical economic interactions.
\cite{bai_evolution_2022} analyze the evolution of user's Ethereum transaction patterns using a temporal graph perspective. They construct different transaction graphs and examine their structural changes over time.
\citet{shen_identity_2021} introduce a novel approach for analyzing user behavior through transaction subgraphs. This perspective changes the identity inference task into a graph classification problem. They suggest a general end-to-end graph neural network model I\(^2\)BGNN. The results of the conducted experiments show that I\(^2\)BGNN outperforms other methods like Graph2vec, SF, NetIsd, FGSD on both EOS and Ethereum datasets.
\citet{lee_measurements_2020} investigate interactions between users and smart contracts on Ethereum. Their experiments reveal that blockchain networks are small and connected, differing significantly from traditional social networks.
\cite{mizerka_role_2020} analyze Bitcoin user transactions through a graph-based approach. They investigate whether Bitcoin’s price is influenced more by financial markets or by the behavior of major Bitcoin users. The study finds that major Bitcoin users significantly impact its price.

\subsection{Community Detection}
Researchers aim to identify related users by recognizing a social network, focusing not on a single user but rather on a group of users and their interactions. The basis of community analysis is typically the transaction network of a blockchain or cryptocurrency. This topic is closely related to the detection of illegal activities, users and user groups, respectively associated malicious users, on a blockchain network, as well as user analysis. A widely used technique to detect communities in blockchain networks is address clustering by analyzing the transaction network of the blockchain. We list the identified 10 key publications for community detection in Table \ref{tab:Key-Paper-Community}.

\begin{table*}[]
\footnotesize
\centering
\begin{tabular} { p{0.05\textwidth}  p{0.25\textwidth} p{0.25\textwidth} p{0.25\textwidth} p{0.09\textwidth} }
\multicolumn{4}{l}{\textbf{Top 10 key publications for Community Detection}}\\
\hline
\textbf{Ref (\#)}  & \textbf{Dataset} & \textbf{Data Processing} & \textbf{Result} & \textbf{Platform} \\ \hline
\cite{victor_address_2020} (44) & Full ETH ledger with focus of transaction and event data & Heuristic clustering algorithms & Detection and clustering of related addresses & ETH \\ \hline
\cite{harrigan_airdrops_2018} (43) & Full transaction log of four blockchains & Address clustering of airdrops across multiple blockchains & Impact measures of sharing addresses across blockchains & BTC, LC, DC, CL \\ \hline
\cite{he_bitcoin_2022} (41) & Addresses with respective transactions & Heuristic clustering algorithms & Clustered addresses with graph visualization for community detection & BTC \\ \hline
\cite{oggier_biva_2018} (31) & Addresses transaction graph & Graph clustering algorithms & Visual clustered transaction graph with related addresses & BTC \\ \hline
\cite{sun_ethereum_2019} (20) & 10 million ETH transactions & Metapath2vec and T-distributed stochastic neighbor embedding & Detection of related addresses & ETH \\ \hline
\cite{chang_improving_2020} (14) & BTC transactions & Heuristic address clustering algorithms & Clustered related addresses with respective cluster transaction patterns & BTC \\ \hline
\cite{lee_machine_2020} (14) & Transaction hash list with associated addresses & Random forest and Artificial Neural Network & Clustered and classified addresses & BTC \\ \hline
\cite{moser_resurrecting_2022} (12) & BTC transactions & Random forest & Clustered addresses with output change prediction & BTC \\ \hline
\cite{alqassem_anti-social_2020} (12) & 11 time-spaced sequential snapshots & Graph analysis measurements & Social network properties of BTC & BTC \\ \hline
\cite{mcginn_towards_2018} (9) & Full BTC ledger & Neo4j for graph construction, graph analysis and visualization & Visual analysis of the full blockchain, coinbase blocks, span and users & BTC \\ \hline
\end{tabular}
\caption{List of the top 10 key publications for the topic cluster Community Detection included in the 55 key publications in Blockchain Data Analytics based on the number of citations (\#) per paper. Legend: BTC - Bitcoin, CL - Clam, DC - Dogecoin, ETH - Ethereum, LC - Litecoin}
\label{tab:Key-Paper-Community}
\end{table*}

\citet{victor_address_2020} introduces similar heuristics for Ethereum, focusing on address reuse, airdrop participation, and token spending authorization. Their analysis shows that address reuse is the most effective clustering heuristic, though exchange service addresses can distort results.
\citet{harrigan_airdrops_2018} explore how cryptocurrency airdrops can compromise user privacy. Their analysis shows that shared addresses across multiple blockchains can reveal ownership, exposing users to tracking risks. The authors urge privacy-conscious users to be cautious with such transactions.

\citet{he_bitcoin_2022} improve Bitcoin community detection by combining six heuristic conditions, therefore uncovering hidden connections between Bitcoin addresses. Their method significantly enhances entity and community identification accuracy.
\citet{oggier_biva_2018} use graph analysis algorithms for wallet address aggregation. Their tool is demonstrated by analyzing blackmail extortion linked to the \textit{Ashley Madison} breach, revealing connections between receiving addresses and other scams.
\citet{sun_ethereum_2019} apply machine learning to analyze Ethereum transaction behavior, clustering related users and smart contracts. The largest identified clusters are centered around exchange markets or malicious activity. This highlights the link between community detection, illegal activity detection and user analysis.
\citet{chang_improving_2020} use heuristic clustering to analyze Bitcoin addresses based on transaction patterns, flagging those matching specific patterns and examining input-output relationships. 

\citet{lee_machine_2020} use machine learning to classify Bitcoin addresses based on transaction behavior. By extracting 80 features and using Random Forest and ANN they achieve a near-perfect precision for darknet-related addresses. Their approach can improve blockchain data analytics, helping authorities detect suspicious user clusters.
\citet{moser_resurrecting_2022} enhance Bitcoin address clustering by evaluating 26 heuristics and applying a random forest classifier to improve change address identification. The study highlights the importance of reliable clustering for law enforcement and regulatory compliance, balancing privacy and transparency.
\citet{alqassem_anti-social_2020} analyze the social network properties of Bitcoin by comparing its transaction graph to networks like Facebook, Twitter, and Google+. Their analysis finds Bitcoin’s longest connected component is similar to those of other social networks, but with a significantly larger diameter.

\citet{mcginn_towards_2018} illustrate Bitcoin's open data benefits by converting its binary structure into a graph model. Their analysis reveals transaction patterns linking user activity, traces wealth accumulation from newly mined bitcoin, measures Bitcoin’s disinflationary traits, and enables user defense against attacks.

\subsection{Mining Analysis}
Mining is crucial for the integrity of blockchain networks with a high degree of centralization and high security requirements. The analysis of mining behavior, pools, and strategies is essential to ensure network health, assess mining profitability and efficiency. This analysis focuses on understanding the structural, behavioral, and dynamic aspects of the mining ecosystem to ensure network security and functionality. It includes studying mining pools, computational capabilities, trends, and vulnerabilities, while also considering economic factors, linking it to the financial analysis of blockchain systems. We list the identified 10 key publications for mining analysis in Table \ref{tab:Key-Paper-Mining}.

\begin{table*}[]
\footnotesize
\centering
\begin{tabular} { p{0.05\textwidth}  p{0.25\textwidth} p{0.25\textwidth} p{0.25\textwidth} p{0.09\textwidth} }
\multicolumn{4}{l}{\textbf{Top 10 key publications for Mining Analysis}}\\
\hline
\textbf{Ref (\#)}  & \textbf{Dataset} & \textbf{Data Processing} & \textbf{Result} & \textbf{Platform} \\ \hline
\cite{tovanich_pattern_2023} (13) & BTC directed acyclic transaction network & Unbiased random walk and shortest path walk & Mining pool identification with mining behavior & BTC \\ \hline
\cite{tovanich_interactive_2021} (5) & Mining pool transactions and properties, BTC news and BTC network statistics & Graph visualization and statistical analysis & Visualization and property analysis of mining pools & BTC \\ \hline
\cite{wang_dissecting_2023} (4) & Full BTC node synchronization & Mining pool modeling, behavior and performance analysis & Mining pool property and performance measures & BTC \\ \hline
\cite{tovanich_miningvis_2022} (3) & Mining pool distribution and characteristics, BTC network statistics, BTC news, miner migration in mining pools & Graph visualization and statistical analysis & Visualization of mining, mining pools and mining statistics related to BTC news & BTC \\ \hline
\cite{niu_incentive_2021} (2) & BTC-NG block mining information & MDP Matlab library for Markov decision analysis & Incentive analysis for miners and mining pools on BTC-NG & BTC-NG \\ \hline
\end{tabular}
\caption{List of the 5 publications in the topic cluster Mining Analysis included in the 55 key publications in Blockchain Data Analytics based on the number of citations (\#) per paper. Legend: BTC - Bitcoin, BTC-NG - Bitcoin-NG}
\label{tab:Key-Paper-Mining}
\end{table*}

\citet{tovanich_pattern_2023} present a method to identify taint flows-dynamic networks that track Bitcoin transfers from an initial source through various recipients until they dissipate. The authors demonstrate the efficacy of their embedding method for classifying mining pools. They show that the method yields high accuracy when used with a supervised learning approach, as opposed to an unsupervised learning approach.
In order to investigate the long-term historical development and dynamics of the Bitcoin mining ecosystem, \citet{tovanich_miningvis_2022} introduce a visual analytics tool that analyzes Bitcoin mining dynamics, including pool rankings, market stats, and pool-hopping patterns. While useful for analysts, its value for miners is limited to historical trend analysis. The tool is further demonstrated in \citet{tovanich_interactive_2021}.
\citet{wang_dissecting_2023} analyze 12 million unconfirmed transactions and 254,000 Bitcoin blocks to study mining pool dynamics. They find that a few pools control most computational power, top pools' short-term mining declines but long-term power grows exponentially, and network computing nears a Nash equilibrium. The study suggests game-theoretic strategies for mining behavior but notes limitations due to its assumption of non-adversarial competition.
\citet{niu_incentive_2021} examine Bitcoin-NG’s incentive mechanisms, addressing gaps in prior studies by considering network capacity and both key/microblock incentives. Using a Markov Decision Process, they find Bitcoin-NG remains incentive-compatible but is vulnerable to selfish mining if an attacker controls over 35\% of mining power. The study offers insights for improving blockchain protocol security through careful system design.

\section{Discussion Results and Future Challenges}
\label{Discussion}

During our analysis, we identified 55 key publications that characterize the domain of Blockchain Data Analytics. These core publications are the ten most cited publications for each of six topic clusters, thus standing for the most relevant contributions in the respective topic cluster. In this section, we answer the research questions based on the derived topic clusters with their respective key publications.

\textit{RQ1: In which outlets is research on blockchain data analytics published and by which institutions?} Research on blockchain data analytics is published in a wide array of outlets and no clearly leading outlets could be identified. The top outlets, illustrated in Table \ref{tab:Top-Outlets}, show that IEEE conferences and journals are on the forefront in blockchain data analytics in terms of the number of publications. Our descriptive analysis further showed that China is the biggest contributor to the research conducted in blockchain data analytics. The United States and India still belong to the top contributors, although they contribute considerably less in absolute values. Mostly the research is well distributed between institutions of a country, while only the Sun Yat-Sen University in China stands out with 24 publications. 


\textit{RQ2: What are the major topics of blockchain data analytics?} Based on an initial set of 466 papers that we consider to represent the core contributions in blockchain data analytics research, we were able to manually derive six topic clusters: \emph{illegal activity detection}, \emph{data management}, \emph{financial analysis}, \emph{user analysis}, \emph{community detection} and \emph{mining analysis}. These inform about the application areas of blockchain data analytics that have been explored in the past. Based on a close examination of the 55 most relevant papers in terms of citations, we found that the applied analysis techniques in these topic clusters are mostly tailored to the analysis goals. 
Throughout all topic clusters, the analysis techniques are often user-centric. Many techniques in illegal activity detection leverage transactions and transaction graph analysis in order to detect certain activities or users. Approaches within the topic clusters of community detection and user analysis search for insights in user behavior, user data and user relationships. The financial analysis cluster often excludes users, focusing more on the financial aspects of a blockchain network, where user-centric analysis is conducted through wealth distribution analysis. Mining analysis is user-centric in examining miners' returns, performance, and mining distribution rather than blockchain users. Data management, with techniques like database and query layers, is not viewed as a data analysis technique itself, but serves as an enabler for data analysis. 

\textit{RQ3: How did the identified topics in blockchain data analytics evolve over time?} As we showed in Section \ref{Review:Topic_Categorization}, the total number of publications in blockchain data analytics increased steadily since 2013. Especially, the topic of illegal activity detection has seen remarkable growth.
One reason may be the rise in blockchain misuse for illegal activities as well as advancements in machine learning that enable innovative applications to blockchain data analytics, which are particularly useful for law enforcement.
Our analysis indicates a substantial increase in interest in blockchain data analytics from 2017 onwards, which has persisted and grown.
Following the initial period of significant hype surrounding cryptocurrencies, the focus then shifted to price prediction approaches. The implementation of data management methodologies, such as query layers or databases operating on a blockchain, has also witnessed a surge in interest since 2017. The field of community detection has not undergone a similar expansion. One reason may be its close association with detecting illegal activities, as many papers in this topic cluster build upon this motivation. 
If the current trend persists, illegal activity detection remains the main focus of blockchain data analytics. However, a first decline in publications appeared between 2022 to 2023.

\textit{RQ4: Which topics in blockchain data analytics research are currently missing in comparison to the general scope of business intelligence?}
A comparison of the state-of-the-art in blockchain data analytics research to the three key perspectives in business intelligence reveals that previous research contributions have predominantly focused on the \emph{production} and \emph{customer} perspectives. The \emph{organizational} perspective, on the other hand, has received minimal attention. 
Blockchain data analytics research is so far mostly product/operations- and user-centric as indicated by the strong focus on illegal behaviors and user aspects such as de-anonymization and transaction properties. 
Organizational topics like Distributed Autonomous Organizations (DAO), business aspects of smart contract behavior, integration with off-chain enterprise data, or the embedding in business processes of organizations are only briefly explored, e.g. \cite{arroyo_dao-analyzer_2022}. Notable exceptions include the approach by \citet{muhlberger_extracting_2019} who demonstrated the potential value of process mining for blockchain systems to enterprises and the contribution by~\citet{klinkmuller_mining_2019} who provide a method for developers to analyze and optimize blockchain-based processes by converting raw transaction data into process mining-ready logs. However, apart from these contributions, the organizational perspective of business intelligence for blockchain is in its infancy.
Blockchain data analytics currently lacks a holistic approach linking all three business intelligence perspectives. The integration of blockchain data analytics into the performance management of enterprises, as previously discussed for non-blockchain applications~\citep{schlafke_framework_2013}, has the potential to enhance decision-making in commercial use cases. Performance management analytics confers a competitive advantage to enterprises by facilitating the identification of market opportunities, threats, and changes. This requires precise and pertinent data for effective decision-making. Recent results in marketing research illustrate this potential, by highlighting how customer relationship management (CRM) may benefit from the transparency of blockchain data~\citep{hanneke_decoding_2024}. 

\textit{RQ5: What datasets, data processing approaches and results are used in blockchain data analytics?} 
Our key publications show that the used datasets are very diverse, but most analysis approaches use transactions as the base dataset, often transforming them into graphs to study users or transactions. Blockchain data is rarely supplemented with off-chain information, e.g. \cite{mizerka_role_2020}.
In cryptocurrency price prediction, blockchain properties are mapped to exchange rates. Machine learning and graph analysis are widely used across topics, especially for illegal activity detection, alongside statistical analysis and heuristic algorithms.

\section{Conclusion}
\label{Conclusion}
This study systematically examined the intersection of blockchain technology and data analytics, thereby revealing the considerable potential that data-driven methodologies hold for enhancing blockchain use cases. A comprehensive literature review was conducted to identify the key research directions, which were subsequently categorized into six clusters. These clusters include the detection of illegal activities, the identification of communities, the analysis of mining, the analysis of user behavior, the analysis of financial data, and the analysis of data management.

Our findings highlight the evolution of blockchain data analytics from simple transaction tracking to sophisticated data integration and machine learning applications. A major area of growth is fraud detection, where advanced anomaly detection models and machine learning techniques help identify illicit activities, reflecting both the increasing sophistication of cybercrime and the growing role of analytics in blockchain security. Community detection and user analysis have also gained prominence, employing graph-based methods to uncover behavioral patterns and network dynamics. Similarly, financial analysis has expanded to include price prediction and investment strategies, emphasizing the economic implications of blockchain data. Furthermore, the study underscores the importance of effective data management techniques, such as query layers and blockchain databases, which enable efficient organization and retrieval of blockchain data. 

Despite these advancements, our analysis uncovered significant gaps in the current research landscape. Notably, there is a lack of holistic frameworks that integrate blockchain metrics with organizational aspects such as key performance indicators (KPIs). A first proposal in this direction has been recently suggested~\citep{curty_blockchain_2025}. Such an approach would bridge the gap between blockchain technology and business intelligence, enabling organizations to derive actionable insights from their blockchain-based operations. This need for integration is particularly salient as an increasing number of companies investigate the potential of blockchain solutions for commercial application.
Future research should focus on unified frameworks integrating on-chain and off-chain data, along with analyzing Decentralized Autonomous Organizations, business behavior of smart contracts, and integrations with enterprise systems.

In conclusion, blockchain data analytics offers significant potential for enhancing transparency, security, and efficiency across a range of sectors. 
This study provides a foundation for future research, emphasizing the need for comprehensive, integrated approaches to blockchain data analytics.

\begin{acks}
This work was partially funded by the Swiss National Science Foundation project Domain-Specific Conceptual Modeling for Distributed Ledger Technologies [196889].
\end{acks}

\bibliographystyle{ACM-Reference-Format}

\bibliography{references_zotero}


\begin{thebibliography}{113}


\ifx \showCODEN    \undefined \def \showCODEN     #1{\unskip}     \fi
\ifx \showDOI      \undefined \def \showDOI       #1{#1}\fi
\ifx \showISBNx    \undefined \def \showISBNx     #1{\unskip}     \fi
\ifx \showISBNxiii \undefined \def \showISBNxiii  #1{\unskip}     \fi
\ifx \showISSN     \undefined \def \showISSN      #1{\unskip}     \fi
\ifx \showLCCN     \undefined \def \showLCCN      #1{\unskip}     \fi
\ifx \shownote     \undefined \def \shownote      #1{#1}          \fi
\ifx \showarticletitle \undefined \def \showarticletitle #1{#1}   \fi
\ifx \showURL      \undefined \def \showURL       {\relax}        \fi
\providecommand\bibfield[2]{#2}
\providecommand\bibinfo[2]{#2}
\providecommand\natexlab[1]{#1}
\providecommand\showeprint[2][]{arXiv:#2}

\bibitem[Akcora et~al\mbox{.}(2018)]%
        {akcora_blockchain_2018}
\bibfield{author}{\bibinfo{person}{Cuneyt~Gurcan Akcora}, \bibinfo{person}{Murat Kantarcioglu}, {and} \bibinfo{person}{Yulia~R. Gel}.} \bibinfo{year}{2018}\natexlab{}.
\newblock \showarticletitle{Blockchain {Data} {Analytics}}. In \bibinfo{booktitle}{\emph{2018 {IEEE} {International} {Conference} on {Data} {Mining} ({ICDM})}}. \bibinfo{pages}{6--6}.
\newblock
\urldef\tempurl%
\url{https://doi.org/10.1109/ICDM.2018.00013}
\showDOI{\tempurl}


\bibitem[Alarab et~al\mbox{.}(2020)]%
        {alarab_competence_2020}
\bibfield{author}{\bibinfo{person}{Ismail Alarab}, \bibinfo{person}{Simant Prakoonwit}, {and} \bibinfo{person}{Mohamed~Ikbal Nacer}.} \bibinfo{year}{2020}\natexlab{}.
\newblock \showarticletitle{Competence of {Graph} {Convolutional} {Networks} for {Anti}-{Money} {Laundering} in {Bitcoin} {Blockchain}}. In \bibinfo{booktitle}{\emph{Proceedings of the 2020 5th {International} {Conference} on {Machine} {Learning} {Technologies}}} \emph{(\bibinfo{series}{{ICMLT} '20})}. \bibinfo{publisher}{Association for Computing Machinery}, \bibinfo{address}{New York, NY, USA}, \bibinfo{pages}{23--27}.
\newblock
\showISBNx{978-1-4503-7764-5}
\urldef\tempurl%
\url{https://doi.org/10.1145/3409073.3409080}
\showDOI{\tempurl}


\bibitem[Alqassem et~al\mbox{.}(2020)]%
        {alqassem_anti-social_2020}
\bibfield{author}{\bibinfo{person}{Israa Alqassem}, \bibinfo{person}{Iyad Rahwan}, {and} \bibinfo{person}{Davor Svetinovic}.} \bibinfo{year}{2020}\natexlab{}.
\newblock \showarticletitle{The {Anti}-{Social} {System} {Properties}: {Bitcoin} {Network} {Data} {Analysis}}.
\newblock \bibinfo{journal}{\emph{IEEE Transactions on Systems, Man, and Cybernetics: Systems}} \bibinfo{volume}{50}, \bibinfo{number}{1} (\bibinfo{year}{2020}), \bibinfo{pages}{21--31}.
\newblock
\urldef\tempurl%
\url{https://doi.org/10.1109/TSMC.2018.2883678}
\showDOI{\tempurl}


\bibitem[Amarasinghe et~al\mbox{.}(2019)]%
        {amarasinghe_survey_2019}
\bibfield{author}{\bibinfo{person}{Niluka Amarasinghe}, \bibinfo{person}{Xavier Boyen}, {and} \bibinfo{person}{Matthew McKague}.} \bibinfo{year}{2019}\natexlab{}.
\newblock \showarticletitle{A {Survey} of {Anonymity} of {Cryptocurrencies}}. In \bibinfo{booktitle}{\emph{Proceedings of the {Australasian} {Computer} {Science} {Week} {Multiconference}}} \emph{(\bibinfo{series}{{ACSW} '19})}. \bibinfo{publisher}{Association for Computing Machinery}, \bibinfo{address}{New York, NY, USA}.
\newblock
\showISBNx{978-1-4503-6603-8}
\urldef\tempurl%
\url{https://doi.org/10.1145/3290688.3290693}
\showDOI{\tempurl}


\bibitem[Amudha(2022)]%
        {amudha_dilated_2022}
\bibfield{author}{\bibinfo{person}{G. Amudha}.} \bibinfo{year}{2022}\natexlab{}.
\newblock \showarticletitle{Dilated {Transaction} {Access} and {Retrieval}: {Improving} the {Information} {Retrieval} of {Blockchain}-{Assimilated} {Internet} of {Things} {Transactions}}.
\newblock \bibinfo{journal}{\emph{Wireless Personal Communications}} \bibinfo{volume}{127}, \bibinfo{number}{1} (\bibinfo{date}{Nov.} \bibinfo{year}{2022}), \bibinfo{pages}{85--105}.
\newblock
\showISSN{1572-834X}
\urldef\tempurl%
\url{https://doi.org/10.1007/s11277-021-08094-y}
\showDOI{\tempurl}


\bibitem[Arroyo et~al\mbox{.}(2022)]%
        {arroyo_dao-analyzer_2022}
\bibfield{author}{\bibinfo{person}{Javier Arroyo}, \bibinfo{person}{David Davó}, \bibinfo{person}{Elena Martínez-Vicente}, \bibinfo{person}{Youssef Faqir-Rhazoui}, {and} \bibinfo{person}{Samer Hassan}.} \bibinfo{year}{2022}\natexlab{}.
\newblock \showarticletitle{{DAO}-{Analyzer}: {Exploring} {Activity} and {Participation} in {Blockchain} {Organizations}}. In \bibinfo{booktitle}{\emph{Companion {Publication} of the 2022 {Conference} on {Computer} {Supported} {Cooperative} {Work} and {Social} {Computing}}} \emph{(\bibinfo{series}{{CSCW}'22 {Companion}})}. \bibinfo{publisher}{Association for Computing Machinery}, \bibinfo{address}{New York, NY, USA}, \bibinfo{pages}{193--196}.
\newblock
\showISBNx{978-1-4503-9190-0}
\urldef\tempurl%
\url{https://doi.org/10.1145/3500868.3559707}
\showDOI{\tempurl}


\bibitem[Bai et~al\mbox{.}(2022)]%
        {bai_evolution_2022}
\bibfield{author}{\bibinfo{person}{Qianlan Bai}, \bibinfo{person}{Chao Zhang}, \bibinfo{person}{Nianyi Liu}, \bibinfo{person}{Xiaowei Chen}, \bibinfo{person}{Yuedong Xu}, {and} \bibinfo{person}{Xin Wang}.} \bibinfo{year}{2022}\natexlab{}.
\newblock \showarticletitle{Evolution of {Transaction} {Pattern} in {Ethereum}: {A} {Temporal} {Graph} {Perspective}}.
\newblock \bibinfo{journal}{\emph{IEEE Transactions on Computational Social Systems}} \bibinfo{volume}{9}, \bibinfo{number}{3} (\bibinfo{year}{2022}), \bibinfo{pages}{851--866}.
\newblock
\urldef\tempurl%
\url{https://doi.org/10.1109/TCSS.2021.3108788}
\showDOI{\tempurl}


\bibitem[Balaskas and Franqueira(2018)]%
        {balaskas_analytical_2018}
\bibfield{author}{\bibinfo{person}{Anastasios Balaskas} {and} \bibinfo{person}{Virginia N.~L. Franqueira}.} \bibinfo{year}{2018}\natexlab{}.
\newblock \showarticletitle{Analytical {Tools} for {Blockchain}: {Review}, {Taxonomy} and {Open} {Challenges}}. In \bibinfo{booktitle}{\emph{2018 {International} {Conference} on {Cyber} {Security} and {Protection} of {Digital} {Services} ({Cyber} {Security})}}. \bibinfo{pages}{1--8}.
\newblock
\urldef\tempurl%
\url{https://doi.org/10.1109/CyberSecPODS.2018.8560672}
\showDOI{\tempurl}


\bibitem[Banafa(2023)]%
        {banafa_introduction_2023}
\bibfield{author}{\bibinfo{person}{Ahmed Banafa}.} \bibinfo{year}{2023}\natexlab{}.
\newblock \bibinfo{booktitle}{\emph{Introduction to {Blockchain} {Technology}}}.
\newblock \bibinfo{publisher}{Taylor and Francies {\textbackslash}textbackslashtextbar CRC Press River Publishers}.
\newblock
\showISBNx{978-87-7022-160-3}


\bibitem[Bandara et~al\mbox{.}(2018)]%
        {bandara_mystikoblockchain_2018}
\bibfield{author}{\bibinfo{person}{Eranga Bandara}, \bibinfo{person}{Wee~Keong Ng}, \bibinfo{person}{Kasun De~Zoysa}, \bibinfo{person}{Fernando Newton}, \bibinfo{person}{Tharaka Hewa}, \bibinfo{person}{P. Maurakirinathan}, {and} \bibinfo{person}{Namal Jayasuriya}.} \bibinfo{year}{2018}\natexlab{}.
\newblock \showarticletitle{Mystiko—{Blockchain} {Meets} {Big} {Data}}. In \bibinfo{booktitle}{\emph{2018 {IEEE} {International} {Conference} on {Big} {Data} ({Big} {Data})}}. \bibinfo{pages}{3024--3032}.
\newblock
\urldef\tempurl%
\url{https://doi.org/10.1109/BigData.2018.8622341}
\showDOI{\tempurl}


\bibitem[Banerjee et~al\mbox{.}(2013)]%
        {banerjee_data_2013}
\bibfield{author}{\bibinfo{person}{Arindam Banerjee}, \bibinfo{person}{Tathagata Bandyopadhyay}, {and} \bibinfo{person}{Prachi Acharya}.} \bibinfo{year}{2013}\natexlab{}.
\newblock \showarticletitle{Data {Analytics}: {Hyped} {Up} {Aspirations} or {True} {Potential}?}
\newblock \bibinfo{journal}{\emph{Vikalpa}} \bibinfo{volume}{38}, \bibinfo{number}{4} (\bibinfo{date}{Oct.} \bibinfo{year}{2013}), \bibinfo{pages}{1--12}.
\newblock
\showISSN{0256-0909}
\urldef\tempurl%
\url{https://doi.org/10.1177/0256090920130401}
\showDOI{\tempurl}


\bibitem[Bartoletti et~al\mbox{.}(2018)]%
        {bartoletti_data_2018}
\bibfield{author}{\bibinfo{person}{Massimo Bartoletti}, \bibinfo{person}{Barbara Pes}, {and} \bibinfo{person}{Sergio Serusi}.} \bibinfo{year}{2018}\natexlab{}.
\newblock \showarticletitle{Data {Mining} for {Detecting} {Bitcoin} {Ponzi} {Schemes}}. In \bibinfo{booktitle}{\emph{2018 {Crypto} {Valley} {Conference} on {Blockchain} {Technology} ({CVCBT})}}. \bibinfo{pages}{75--84}.
\newblock
\urldef\tempurl%
\url{https://doi.org/10.1109/CVCBT.2018.00014}
\showDOI{\tempurl}


\bibitem[Bartoletti and Pompianu(2017)]%
        {bartoletti_analysis_2017}
\bibfield{author}{\bibinfo{person}{Massimo Bartoletti} {and} \bibinfo{person}{Livio Pompianu}.} \bibinfo{year}{2017}\natexlab{}.
\newblock \showarticletitle{An {Analysis} of {Bitcoin} {OP}\_RETURN {Metadata}}. In \bibinfo{booktitle}{\emph{Financial {Cryptography} and {Data} {Security}}}, \bibfield{editor}{\bibinfo{person}{Michael Brenner}, \bibinfo{person}{Kurt Rohloff}, \bibinfo{person}{Joseph Bonneau}, \bibinfo{person}{Andrew Miller}, \bibinfo{person}{Peter~Y.A. Ryan}, \bibinfo{person}{Vanessa Teague}, \bibinfo{person}{Andrea Bracciali}, \bibinfo{person}{Massimiliano Sala}, \bibinfo{person}{Federico Pintore}, {and} \bibinfo{person}{Markus Jakobsson}} (Eds.). \bibinfo{publisher}{Springer International Publishing}, \bibinfo{address}{Cham}, \bibinfo{pages}{218--230}.
\newblock
\showISBNx{978-3-319-70278-0}


\bibitem[Bergman and Rajput(2021)]%
        {bergman_revealing_2021}
\bibfield{author}{\bibinfo{person}{Karolina Bergman} {and} \bibinfo{person}{Saeed Rajput}.} \bibinfo{year}{2021}\natexlab{}.
\newblock \showarticletitle{Revealing and {Concealing} {Bitcoin} {Identities}: {A} {Survey} of {Techniques}}. In \bibinfo{booktitle}{\emph{Proceedings of the 3rd {ACM} {International} {Symposium} on {Blockchain} and {Secure} {Critical} {Infrastructure}}} \emph{(\bibinfo{series}{{BSCI} '21})}. \bibinfo{publisher}{Association for Computing Machinery}, \bibinfo{address}{New York, NY, USA}, \bibinfo{pages}{13--24}.
\newblock
\showISBNx{978-1-4503-8400-1}
\urldef\tempurl%
\url{https://doi.org/10.1145/3457337.3457838}
\showDOI{\tempurl}


\bibitem[Biryukov and Tikhomirov(2019)]%
        {biryukov_deanonymization_2019}
\bibfield{author}{\bibinfo{person}{Alex Biryukov} {and} \bibinfo{person}{Sergei Tikhomirov}.} \bibinfo{year}{2019}\natexlab{}.
\newblock \showarticletitle{Deanonymization and {Linkability} of {Cryptocurrency} {Transactions} {Based} on {Network} {Analysis}}. In \bibinfo{booktitle}{\emph{2019 {IEEE} {European} {Symposium} on {Security} and {Privacy} ({EuroS}\&{P})}}. \bibinfo{pages}{172--184}.
\newblock
\urldef\tempurl%
\url{https://doi.org/10.1109/EuroSP.2019.00022}
\showDOI{\tempurl}


\bibitem[Boell and Cecez-Kecmanov(2012)]%
        {boell_conceptualizing_2012}
\bibfield{author}{\bibinfo{person}{Sebastian Boell} {and} \bibinfo{person}{Dubravka Cecez-Kecmanov}.} \bibinfo{year}{2012}\natexlab{}.
\newblock \showarticletitle{Conceptualizing {Information} {Systems}: {From} '{Input}-{Processing}-{Output}' {Devices} {To} {Sociomaterial} {Apparatuses}}.
\newblock \bibinfo{journal}{\emph{ECIS 2012 Proceedings}} (\bibinfo{date}{May} \bibinfo{year}{2012}).
\newblock
\urldef\tempurl%
\url{https://aisel.aisnet.org/ecis2012/20}
\showURL{%
\tempurl}


\bibitem[Brainard(2024)]%
        {brainard_fast-growing_2024}
\bibfield{author}{\bibinfo{person}{Jeffrey Brainard}.} \bibinfo{year}{2024}\natexlab{}.
\newblock \bibinfo{title}{Fast-growing open-access journals stripped of coveted impact factors}.
\newblock
\newblock
\urldef\tempurl%
\url{https://www.science.org/content/article/fast-growing-open-access-journals-stripped-coveted-impact-factors}
\showURL{%
\tempurl}


\bibitem[Chan and Olmsted(2017)]%
        {chan_ethereum_2017}
\bibfield{author}{\bibinfo{person}{Wren Chan} {and} \bibinfo{person}{Aspen Olmsted}.} \bibinfo{year}{2017}\natexlab{}.
\newblock \showarticletitle{Ethereum transaction graph analysis}. In \bibinfo{booktitle}{\emph{2017 12th {International} {Conference} for {Internet} {Technology} and {Secured} {Transactions} ({ICITST})}}. \bibinfo{pages}{498--500}.
\newblock
\urldef\tempurl%
\url{https://doi.org/10.23919/ICITST.2017.8356459}
\showDOI{\tempurl}


\bibitem[Chang and Svetinovic(2020)]%
        {chang_improving_2020}
\bibfield{author}{\bibinfo{person}{Tao-Hung Chang} {and} \bibinfo{person}{Davor Svetinovic}.} \bibinfo{year}{2020}\natexlab{}.
\newblock \showarticletitle{Improving {Bitcoin} {Ownership} {Identification} {Using} {Transaction} {Patterns} {Analysis}}.
\newblock \bibinfo{journal}{\emph{IEEE Transactions on Systems, Man, and Cybernetics: Systems}} \bibinfo{volume}{50}, \bibinfo{number}{1} (\bibinfo{year}{2020}), \bibinfo{pages}{9--20}.
\newblock
\urldef\tempurl%
\url{https://doi.org/10.1109/TSMC.2018.2867497}
\showDOI{\tempurl}


\bibitem[Chen et~al\mbox{.}(2020a)]%
        {chen_phishing_2020}
\bibfield{author}{\bibinfo{person}{Weili Chen}, \bibinfo{person}{Xiongfeng Guo}, \bibinfo{person}{Zhiguang Chen}, \bibinfo{person}{Zibin Zheng}, {and} \bibinfo{person}{Yutong Lu}.} \bibinfo{year}{2020}\natexlab{a}.
\newblock \showarticletitle{Phishing {Scam} {Detection} on {Ethereum}: {Towards} {Financial} {Security} for {Blockchain} {Ecosystem}}. In \bibinfo{booktitle}{\emph{Proceedings of the {Twenty}-{Ninth} {International} {Joint} {Conference} on {Artificial} {Intelligence}, {IJCAI}-20}}, \bibfield{editor}{\bibinfo{person}{Christian Bessiere}} (Ed.). \bibinfo{publisher}{International Joint Conferences on Artificial Intelligence Organization}, \bibinfo{pages}{4506--4512}.
\newblock
\urldef\tempurl%
\url{https://doi.org/10.24963/ijcai.2020/621}
\showDOI{\tempurl}


\bibitem[Chen et~al\mbox{.}(2020b)]%
        {chen_traveling_2020}
\bibfield{author}{\bibinfo{person}{Weili Chen}, \bibinfo{person}{Tuo Zhang}, \bibinfo{person}{Zhiguang Chen}, \bibinfo{person}{Zibin Zheng}, {and} \bibinfo{person}{Yutong Lu}.} \bibinfo{year}{2020}\natexlab{b}.
\newblock \showarticletitle{Traveling the token world: {A} graph analysis of {Ethereum} {ERC20} token ecosystem}. In \bibinfo{booktitle}{\emph{Proceedings of {The} {Web} {Conference} 2020}} \emph{(\bibinfo{series}{{WWW} '20})}. \bibinfo{publisher}{Association for Computing Machinery}, \bibinfo{address}{New York, NY, USA}, \bibinfo{pages}{1411--1421}.
\newblock
\showISBNx{978-1-4503-7023-3}
\urldef\tempurl%
\url{https://doi.org/10.1145/3366423.3380215}
\showDOI{\tempurl}


\bibitem[Choubey et~al\mbox{.}(2019)]%
        {choubey_energytradingrank_2019}
\bibfield{author}{\bibinfo{person}{Anurag Choubey}, \bibinfo{person}{Sourajit Behera}, \bibinfo{person}{Yashwant~Singh Patel}, \bibinfo{person}{Karanam Mahidhar}, {and} \bibinfo{person}{Rajiv Misra}.} \bibinfo{year}{2019}\natexlab{}.
\newblock \showarticletitle{{EnergyTradingRank} {Algorithm} for {Truthful} {Auctions} among {EVs} via {Blockchain} {Analytics} of {Large} {Scale} {Transaction} {Graphs}}. In \bibinfo{booktitle}{\emph{2019 11th {International} {Conference} on {Communication} {Systems} \& {Networks} ({COMSNETS})}}. \bibinfo{pages}{1--6}.
\newblock
\urldef\tempurl%
\url{https://doi.org/10.1109/COMSNETS.2019.8711249}
\showDOI{\tempurl}


\bibitem[Cunha et~al\mbox{.}(2022)]%
        {cunha_blockchain_2022}
\bibfield{author}{\bibinfo{person}{João Cunha}, \bibinfo{person}{Ricardo Duarte}, \bibinfo{person}{Tiago Guimarães}, \bibinfo{person}{César Quintas}, {and} \bibinfo{person}{Manuel~Filipe Santos}.} \bibinfo{year}{2022}\natexlab{}.
\newblock \showarticletitle{Blockchain analytics in healthcare: {An} {Overview}}.
\newblock \bibinfo{journal}{\emph{Procedia Computer Science}}  \bibinfo{volume}{201} (\bibinfo{year}{2022}), \bibinfo{pages}{708--713}.
\newblock
\showISSN{1877-0509}
\urldef\tempurl%
\url{https://doi.org/10.1016/j.procs.2022.03.095}
\showDOI{\tempurl}


\bibitem[Curty et~al\mbox{.}(2025)]%
        {curty_blockchain_2025}
\bibfield{author}{\bibinfo{person}{Simon Curty}, \bibinfo{person}{Marcel Bühlmann}, \bibinfo{person}{Hans-Georg Fill}, {and} \bibinfo{person}{Horst Treiblmaier}.} \bibinfo{year}{2025}\natexlab{}.
\newblock \showarticletitle{The {Blockchain} {Balanced} {Scorecard}: {An} {Innovative} {Approach} to {Facilitate} {Strategic} {Management} in {Enterprises}}. In \bibinfo{booktitle}{\emph{Proceedings of the 58th {Hawaii} {International} {Conference} on {System} {Sciences}}}.
\newblock
\showISBNx{978-0-9981331-8-8}
\urldef\tempurl%
\url{https://doi.org/10125/109518}
\showDOI{\tempurl}


\bibitem[Curty and Fill(2023)]%
        {curty_domain-specific_2023}
\bibfield{author}{\bibinfo{person}{Simon Curty} {and} \bibinfo{person}{Hans-Georg Fill}.} \bibinfo{year}{2023}\natexlab{}.
\newblock \showarticletitle{A {Domain}-{Specific} e{\textbackslash}({\textasciicircum}{\textbackslash}mbox3{\textbackslash})value {Extension} for {Analyzing} {Blockchain}-{Based} {Value} {Networks}}. In \bibinfo{booktitle}{\emph{The {Practice} of {Enterprise} {Modeling} - 16th {IFIP} {Working} {Conference}, {PoEM} 2023, {Vienna}, {Austria}, {November} 28 - {December} 1, 2023, {Proceedings}}} \emph{(\bibinfo{series}{Lecture {Notes} in {Business} {Information} {Processing}}, Vol.~\bibinfo{volume}{497})}, \bibfield{editor}{\bibinfo{person}{João Paulo~A. Almeida}, \bibinfo{person}{Monika Kaczmarek-Heß}, \bibinfo{person}{Agnes Koschmider}, {and} \bibinfo{person}{Henderik~A. Proper}} (Eds.). \bibinfo{publisher}{Springer}, \bibinfo{pages}{74--90}.
\newblock
\urldef\tempurl%
\url{https://doi.org/10.1007/978-3-031-48583-1_5}
\showDOI{\tempurl}


\bibitem[Curty et~al\mbox{.}(2023)]%
        {curty_design_2023}
\bibfield{author}{\bibinfo{person}{Simon Curty}, \bibinfo{person}{Felix Härer}, {and} \bibinfo{person}{Hans-Georg Fill}.} \bibinfo{year}{2023}\natexlab{}.
\newblock \showarticletitle{Design of blockchain-based applications using model-driven engineering and low-code/no-code platforms: a structured literature review}.
\newblock \bibinfo{journal}{\emph{Software and Systems Modeling}} \bibinfo{volume}{22}, \bibinfo{number}{6} (\bibinfo{date}{Dec.} \bibinfo{year}{2023}), \bibinfo{pages}{1857--1895}.
\newblock
\showISSN{1619-1374}
\urldef\tempurl%
\url{https://doi.org/10.1007/s10270-023-01109-1}
\showDOI{\tempurl}


\bibitem[Daudt et~al\mbox{.}(2013)]%
        {daudt_enhancing_2013}
\bibfield{author}{\bibinfo{person}{H.M. Daudt}, \bibinfo{person}{C. van Mossel}, {and} \bibinfo{person}{S.J. Scott}.} \bibinfo{year}{2013}\natexlab{}.
\newblock \showarticletitle{Enhancing the scoping study methodology: a large, inter-professional team’s experience with {Arksey} and {O}’{Malley}’s framework {\textbackslash}textbackslashtextbar {BMC} {Medical} {Research} {Methodology} {\textbackslash}textbackslashtextbar {Full} {Text}}.
\newblock \bibinfo{journal}{\emph{BMC Med Res Methodol}} (\bibinfo{year}{2013}).
\newblock
\urldef\tempurl%
\url{https://doi.org/10.1186/1471-2288-13-48}
\showURL{%
\tempurl}
\newblock
\shownote{Edition: 13 Section: 48}.


\bibitem[Di~Francesco~Maesa et~al\mbox{.}(2017a)]%
        {di_francesco_maesa_analysis_2017}
\bibfield{author}{\bibinfo{person}{Damiano Di~Francesco~Maesa}, \bibinfo{person}{Andrea Marino}, {and} \bibinfo{person}{Laura Ricci}.} \bibinfo{year}{2017}\natexlab{a}.
\newblock \showarticletitle{An analysis of the {Bitcoin} users graph: inferring unusual behaviours}. In \bibinfo{booktitle}{\emph{Complex {Networks} \& {Their} {Applications} {V}}}, \bibfield{editor}{\bibinfo{person}{Hocine Cherifi}, \bibinfo{person}{Sabrina Gaito}, \bibinfo{person}{Walter Quattrociocchi}, {and} \bibinfo{person}{Alessandra Sala}} (Eds.). \bibinfo{publisher}{Springer International Publishing}, \bibinfo{address}{Cham}, \bibinfo{pages}{749--760}.
\newblock
\showISBNx{978-3-319-50901-3}


\bibitem[Di~Francesco~Maesa et~al\mbox{.}(2017b)]%
        {di_francesco_maesa_detecting_2017}
\bibfield{author}{\bibinfo{person}{Damiano Di~Francesco~Maesa}, \bibinfo{person}{Andrea Marino}, {and} \bibinfo{person}{Laura Ricci}.} \bibinfo{year}{2017}\natexlab{b}.
\newblock \showarticletitle{Detecting artificial behaviours in the {Bitcoin} users graph}.
\newblock \bibinfo{journal}{\emph{Online Social Networks and Media}}  \bibinfo{volume}{3-4} (\bibinfo{year}{2017}), \bibinfo{pages}{63--74}.
\newblock
\showISSN{2468-6964}
\urldef\tempurl%
\url{https://doi.org/10.1016/j.osnem.2017.10.006}
\showDOI{\tempurl}


\bibitem[Di~Francesco~Maesa et~al\mbox{.}(2018)]%
        {di_francesco_maesa_data-driven_2018}
\bibfield{author}{\bibinfo{person}{Damiano Di~Francesco~Maesa}, \bibinfo{person}{Andrea Marino}, {and} \bibinfo{person}{Laura Ricci}.} \bibinfo{year}{2018}\natexlab{}.
\newblock \showarticletitle{Data-driven analysis of {Bitcoin} properties: exploiting the users graph}.
\newblock \bibinfo{journal}{\emph{International Journal of Data Science and Analytics}} \bibinfo{volume}{6}, \bibinfo{number}{1} (\bibinfo{date}{Aug.} \bibinfo{year}{2018}), \bibinfo{pages}{63--80}.
\newblock
\showISSN{2364-4168}
\urldef\tempurl%
\url{https://doi.org/10.1007/s41060-017-0074-x}
\showDOI{\tempurl}


\bibitem[Dillenberger et~al\mbox{.}(2019)]%
        {dillenberger_blockchain_2019}
\bibfield{author}{\bibinfo{person}{D.~N. Dillenberger}, \bibinfo{person}{P. Novotny}, \bibinfo{person}{Q. Zhang}, \bibinfo{person}{P. Jayachandran}, \bibinfo{person}{H. Gupta}, \bibinfo{person}{S. Hans}, \bibinfo{person}{D. Verma}, \bibinfo{person}{S. Chakraborty}, \bibinfo{person}{J.~J. Thomas}, \bibinfo{person}{M.~M. Walli}, \bibinfo{person}{R. Vaculin}, {and} \bibinfo{person}{K. Sarpatwar}.} \bibinfo{year}{2019}\natexlab{}.
\newblock \showarticletitle{Blockchain analytics and artificial intelligence}.
\newblock \bibinfo{journal}{\emph{IBM Journal of Research and Development}} \bibinfo{volume}{63}, \bibinfo{number}{2/3} (\bibinfo{year}{2019}), \bibinfo{pages}{5:1--5:14}.
\newblock
\urldef\tempurl%
\url{https://doi.org/10.1147/JRD.2019.2900638}
\showDOI{\tempurl}


\bibitem[Dustdar et~al\mbox{.}(2021)]%
        {dustdar_elastic_2021}
\bibfield{author}{\bibinfo{person}{Schahram Dustdar}, \bibinfo{person}{Pablo Fernández}, \bibinfo{person}{José~María García}, {and} \bibinfo{person}{Antonio Ruiz-Cortés}.} \bibinfo{year}{2021}\natexlab{}.
\newblock \showarticletitle{Elastic {Smart} {Contracts} in {Blockchains}}.
\newblock \bibinfo{journal}{\emph{IEEE/CAA Journal of Automatica Sinica}} \bibinfo{volume}{8}, \bibinfo{number}{12} (\bibinfo{year}{2021}), \bibinfo{pages}{1901--1912}.
\newblock
\urldef\tempurl%
\url{https://doi.org/10.1109/JAS.2021.1004222}
\showDOI{\tempurl}


\bibitem[El-Hindi et~al\mbox{.}(2019)]%
        {el-hindi_blockchaindb_2019}
\bibfield{author}{\bibinfo{person}{Muhammad El-Hindi}, \bibinfo{person}{Carsten Binnig}, \bibinfo{person}{Arvind Arasu}, \bibinfo{person}{Donald Kossmann}, {and} \bibinfo{person}{Ravi Ramamurthy}.} \bibinfo{year}{2019}\natexlab{}.
\newblock \showarticletitle{{BlockchainDB}: a shared database on blockchains}.
\newblock \bibinfo{journal}{\emph{Proc. VLDB Endow.}} \bibinfo{volume}{12}, \bibinfo{number}{11} (\bibinfo{date}{July} \bibinfo{year}{2019}), \bibinfo{pages}{1597--1609}.
\newblock
\showISSN{2150-8097}
\urldef\tempurl%
\url{https://doi.org/10.14778/3342263.3342636}
\showDOI{\tempurl}


\bibitem[Elsheikh(2022)]%
        {elsheikh_blockchain_2022}
\bibfield{author}{\bibinfo{person}{Ahmed~S. Elsheikh}.} \bibinfo{year}{2022}\natexlab{}.
\newblock \showarticletitle{Blockchain {Analytics} {Reference} {Architecture} for {FinTech} - {A} {Positioning} {Paper}: {Advancing} {FinTech} with {Blockchain}, {Data} {Analytics}, and {Enterprise} {Architecture}}. In \bibinfo{booktitle}{\emph{Proceedings of the {Federated} {Africa} and {Middle} {East} {Conference} on {Software} {Engineering}}} \emph{(\bibinfo{series}{{FAMECSE} '22})}. \bibinfo{publisher}{Association for Computing Machinery}, \bibinfo{address}{New York, NY, USA}, \bibinfo{pages}{1--7}.
\newblock
\showISBNx{978-1-4503-9663-9}
\urldef\tempurl%
\url{https://doi.org/10.1145/3531056.3531068}
\showDOI{\tempurl}


\bibitem[Farrugia et~al\mbox{.}(2020)]%
        {farrugia_detection_2020}
\bibfield{author}{\bibinfo{person}{Steven Farrugia}, \bibinfo{person}{Joshua Ellul}, {and} \bibinfo{person}{George Azzopardi}.} \bibinfo{year}{2020}\natexlab{}.
\newblock \showarticletitle{Detection of illicit accounts over the {Ethereum} blockchain}.
\newblock \bibinfo{journal}{\emph{Expert Systems with Applications}}  \bibinfo{volume}{150} (\bibinfo{year}{2020}), \bibinfo{pages}{113318}.
\newblock
\showISSN{0957-4174}
\urldef\tempurl%
\url{https://doi.org/10.1016/j.eswa.2020.113318}
\showDOI{\tempurl}


\bibitem[Fill and Meier(2019)]%
        {fill_blockchain_2019}
\bibfield{author}{\bibinfo{person}{Hans-Georg Fill} {and} \bibinfo{person}{Andreas Meier}.} \bibinfo{year}{2019}\natexlab{}.
\newblock \bibinfo{booktitle}{\emph{Blockchain kompakt}}.
\newblock \bibinfo{publisher}{Springer Vieweg Wiesbaden}.
\newblock
\showISBNx{978-3-658-27460-3}


\bibitem[Grossmann and Rinderle-Ma(2015)]%
        {grossmann_fundamentals_2015}
\bibfield{author}{\bibinfo{person}{Wilfried Grossmann} {and} \bibinfo{person}{Stefanie Rinderle-Ma}.} \bibinfo{year}{2015}\natexlab{}.
\newblock \bibinfo{booktitle}{\emph{Fundamentals of {Business} {Intelligence}}}.
\newblock \bibinfo{publisher}{Springer}, \bibinfo{address}{Berlin, Heidelberg}.
\newblock
\showISBNx{978-3-662-46530-1 978-3-662-46531-8}
\urldef\tempurl%
\url{https://doi.org/10.1007/978-3-662-46531-8}
\showDOI{\tempurl}


\bibitem[Han et~al\mbox{.}(2021)]%
        {han_blockchain_2021}
\bibfield{author}{\bibinfo{person}{HuaLong Han}, \bibinfo{person}{YuPeng Chen}, \bibinfo{person}{ChenYing Guo}, {and} \bibinfo{person}{Yin Zhang}.} \bibinfo{year}{2021}\natexlab{}.
\newblock \showarticletitle{Blockchain {Abnormal} {Transaction} {Behavior} {Analysis}: a {Survey}}. In \bibinfo{booktitle}{\emph{Blockchain and {Trustworthy} {Systems}}}, \bibfield{editor}{\bibinfo{person}{Hong-Ning Dai}, \bibinfo{person}{Xuanzhe Liu}, \bibinfo{person}{Daniel~Xiapu Luo}, \bibinfo{person}{Jiang Xiao}, {and} \bibinfo{person}{Xiangping Chen}} (Eds.). \bibinfo{publisher}{Springer Singapore}, \bibinfo{address}{Singapore}, \bibinfo{pages}{57--69}.
\newblock
\showISBNx{978-981-16-7993-3}


\bibitem[Hanneke et~al\mbox{.}(2024)]%
        {hanneke_decoding_2024}
\bibfield{author}{\bibinfo{person}{Björn Hanneke}, \bibinfo{person}{Bernd Skiera}, \bibinfo{person}{Thilo~Gerwien Kraft}, {and} \bibinfo{person}{Oliver Hinz}.} \bibinfo{year}{2024}\natexlab{}.
\newblock \showarticletitle{Decoding blockchain data for research in marketing: {New} insights through an analysis of share of wallet}.
\newblock \bibinfo{journal}{\emph{International Journal of Research in Marketing}} (\bibinfo{date}{Dec.} \bibinfo{year}{2024}).
\newblock
\showISSN{0167-8116}
\urldef\tempurl%
\url{https://doi.org/10.1016/j.ijresmar.2024.12.002}
\showDOI{\tempurl}


\bibitem[Hansen(2024)]%
        {hansen_eliminating_2024}
\bibfield{author}{\bibinfo{person}{Erik~G. Hansen}.} \bibinfo{year}{2024}\natexlab{}.
\newblock \bibinfo{title}{Eliminating {MDPI} journals from publication rating}.
\newblock
\newblock
\urldef\tempurl%
\url{https://www.linkedin.com/posts/prof-erik-g-hansen_publishing-predatorypublishing-openaccess-activity-7251449169691774977-3KJL?utm_source=share&utm_medium=member_desktop}
\showURL{%
\tempurl}


\bibitem[Haque~Sazu and Akter~Jahan(2022)]%
        {haque_sazu_can_2022}
\bibfield{author}{\bibinfo{person}{Mesbaul Haque~Sazu} {and} \bibinfo{person}{Sakila Akter~Jahan}.} \bibinfo{year}{2022}\natexlab{}.
\newblock \showarticletitle{Can big data analytics improve the quality of decision-making in businesses?}
\newblock \bibinfo{journal}{\emph{Iberoamerican Business Journal}} \bibinfo{volume}{6}, \bibinfo{number}{1} (\bibinfo{date}{July} \bibinfo{year}{2022}), \bibinfo{pages}{04--27}.
\newblock
\urldef\tempurl%
\url{https://doi.org/10.22451/5817.ibj2022.vol6.1.11063}
\showDOI{\tempurl}
\newblock
\shownote{Section: NEGOCIOS INTERNACIONALES}.


\bibitem[Harrigan et~al\mbox{.}(2018)]%
        {harrigan_airdrops_2018}
\bibfield{author}{\bibinfo{person}{Martin Harrigan}, \bibinfo{person}{Lei Shi}, {and} \bibinfo{person}{Jacob Illum}.} \bibinfo{year}{2018}\natexlab{}.
\newblock \showarticletitle{Airdrops and {Privacy}: {A} {Case} {Study} in {Cross}-{Blockchain} {Analysis}}. In \bibinfo{booktitle}{\emph{2018 {IEEE} {International} {Conference} on {Data} {Mining} {Workshops} ({ICDMW})}}. \bibinfo{pages}{63--70}.
\newblock
\urldef\tempurl%
\url{https://doi.org/10.1109/ICDMW.2018.00017}
\showDOI{\tempurl}


\bibitem[He et~al\mbox{.}(2022)]%
        {he_bitcoin_2022}
\bibfield{author}{\bibinfo{person}{Xi He}, \bibinfo{person}{Ketai He}, \bibinfo{person}{Shenwen Lin}, \bibinfo{person}{Jinglin Yang}, {and} \bibinfo{person}{Hongliang Mao}.} \bibinfo{year}{2022}\natexlab{}.
\newblock \showarticletitle{Bitcoin address clustering method based on multiple heuristic conditions}.
\newblock \bibinfo{journal}{\emph{IET Blockchain}} \bibinfo{volume}{2}, \bibinfo{number}{2} (\bibinfo{year}{2022}), \bibinfo{pages}{44--56}.
\newblock
\urldef\tempurl%
\url{https://doi.org/10.1049/blc2.12014}
\showDOI{\tempurl}


\bibitem[Hou et~al\mbox{.}(2021)]%
        {hou_survey_2021}
\bibfield{author}{\bibinfo{person}{Wenhan Hou}, \bibinfo{person}{Bo Cui}, {and} \bibinfo{person}{Ru Li}.} \bibinfo{year}{2021}\natexlab{}.
\newblock \showarticletitle{A {Survey} on {Blockchain} {Data} {Analysis}}. In \bibinfo{booktitle}{\emph{2021 {IEEE} 45th {Annual} {Computers}, {Software}, and {Applications} {Conference} ({COMPSAC})}}. \bibinfo{pages}{357--365}.
\newblock
\urldef\tempurl%
\url{https://doi.org/10.1109/COMPSAC51774.2021.00058}
\showDOI{\tempurl}


\bibitem[Hruschka et~al\mbox{.}(2004)]%
        {hruschka_reliability_2004}
\bibfield{author}{\bibinfo{person}{Daniel~J. Hruschka}, \bibinfo{person}{Deborah Schwartz}, \bibinfo{person}{Daphne~Cobb St.John}, \bibinfo{person}{Erin Picone-Decaro}, \bibinfo{person}{Richard~A. Jenkins}, {and} \bibinfo{person}{James~W. Carey}.} \bibinfo{year}{2004}\natexlab{}.
\newblock \showarticletitle{Reliability in {Coding} {Open}-{Ended} {Data}: {Lessons} {Learned} from {HIV} {Behavioral} {Research}}.
\newblock \bibinfo{journal}{\emph{Field Methods}} \bibinfo{volume}{16}, \bibinfo{number}{3} (\bibinfo{date}{Aug.} \bibinfo{year}{2004}), \bibinfo{pages}{307--331}.
\newblock
\showISSN{1525-822X}
\urldef\tempurl%
\url{https://doi.org/10.1177/1525822X04266540}
\showDOI{\tempurl}


\bibitem[Hu et~al\mbox{.}(2021)]%
        {hu_transaction-based_2021}
\bibfield{author}{\bibinfo{person}{Teng Hu}, \bibinfo{person}{Xiaolei Liu}, \bibinfo{person}{Ting Chen}, \bibinfo{person}{Xiaosong Zhang}, \bibinfo{person}{Xiaoming Huang}, \bibinfo{person}{Weina Niu}, \bibinfo{person}{Jiazhong Lu}, \bibinfo{person}{Kun Zhou}, {and} \bibinfo{person}{Yuan Liu}.} \bibinfo{year}{2021}\natexlab{}.
\newblock \showarticletitle{Transaction-based classification and detection approach for {Ethereum} smart contract}.
\newblock \bibinfo{journal}{\emph{Information Processing \& Management}} \bibinfo{volume}{58}, \bibinfo{number}{2} (\bibinfo{year}{2021}), \bibinfo{pages}{102462}.
\newblock
\showISSN{0306-4573}
\urldef\tempurl%
\url{https://doi.org/10.1016/j.ipm.2020.102462}
\showDOI{\tempurl}


\bibitem[Huang et~al\mbox{.}(2021)]%
        {huang_survey_2021}
\bibfield{author}{\bibinfo{person}{Huawei Huang}, \bibinfo{person}{Wei Kong}, \bibinfo{person}{Sicong Zhou}, \bibinfo{person}{Zibin Zheng}, {and} \bibinfo{person}{Song Guo}.} \bibinfo{year}{2021}\natexlab{}.
\newblock \showarticletitle{A {Survey} of {State}-of-the-{Art} on {Blockchains}: {Theories}, {Modelings}, and {Tools}}.
\newblock \bibinfo{journal}{\emph{ACM Comput. Surv.}} \bibinfo{volume}{54}, \bibinfo{number}{2} (\bibinfo{date}{March} \bibinfo{year}{2021}).
\newblock
\showISSN{0360-0300}
\urldef\tempurl%
\url{https://doi.org/10.1145/3441692}
\showDOI{\tempurl}


\bibitem[Jiang and Liu(2021)]%
        {jiang_cryptokitties_2021}
\bibfield{author}{\bibinfo{person}{Xin-Jian Jiang} {and} \bibinfo{person}{Xiao~Fan Liu}.} \bibinfo{year}{2021}\natexlab{}.
\newblock \showarticletitle{{CryptoKitties} {Transaction} {Network} {Analysis}: {The} {Rise} and {Fall} of the {First} {Blockchain} {Game} {Mania}}.
\newblock \bibinfo{journal}{\emph{Frontiers in Physics}}  \bibinfo{volume}{9} (\bibinfo{year}{2021}).
\newblock
\showISSN{2296-424X}
\urldef\tempurl%
\url{https://doi.org/10.3389/fphy.2021.631665}
\showDOI{\tempurl}


\bibitem[Jourdan et~al\mbox{.}(2018)]%
        {jourdan_characterizing_2018}
\bibfield{author}{\bibinfo{person}{Marc Jourdan}, \bibinfo{person}{Sebastien Blandin}, \bibinfo{person}{Laura Wynter}, {and} \bibinfo{person}{Pralhad Deshpande}.} \bibinfo{year}{2018}\natexlab{}.
\newblock \showarticletitle{Characterizing {Entities} in the {Bitcoin} {Blockchain}}. In \bibinfo{booktitle}{\emph{2018 {IEEE} {International} {Conference} on {Data} {Mining} {Workshops} ({ICDMW})}}. \bibinfo{pages}{55--62}.
\newblock
\urldef\tempurl%
\url{https://doi.org/10.1109/ICDMW.2018.00016}
\showDOI{\tempurl}


\bibitem[Kebande et~al\mbox{.}(2022)]%
        {kebande_review_2022}
\bibfield{author}{\bibinfo{person}{Victor~R. Kebande}, \bibinfo{person}{Richard~A. Ikuesan}, {and} \bibinfo{person}{Nickson~M. Karie}.} \bibinfo{year}{2022}\natexlab{}.
\newblock \showarticletitle{Review of {Blockchain} {Forensics} {Challenges}}.
\newblock In \bibinfo{booktitle}{\emph{Blockchain {Security} in {Cloud} {Computing}}}, \bibfield{editor}{\bibinfo{person}{K.M. Baalamurugan}, \bibinfo{person}{S.~Rakesh Kumar}, \bibinfo{person}{Abhishek Kumar}, \bibinfo{person}{Vishal Kumar}, {and} \bibinfo{person}{Sanjeevikumar Padmanaban}} (Eds.). \bibinfo{publisher}{Springer International Publishing}, \bibinfo{address}{Cham}, \bibinfo{pages}{33--50}.
\newblock
\showISBNx{978-3-030-70501-5}
\urldef\tempurl%
\url{https://doi.org/10.1007/978-3-030-70501-5_3}
\showDOI{\tempurl}


\bibitem[Khan(2022)]%
        {khan_graph_2022}
\bibfield{author}{\bibinfo{person}{Arijit Khan}.} \bibinfo{year}{2022}\natexlab{}.
\newblock \showarticletitle{Graph {Analysis} of the {Ethereum} {Blockchain} {Data}: {A} {Survey} of {Datasets}, {Methods}, and {Future} {Work}}. In \bibinfo{booktitle}{\emph{2022 {IEEE} {International} {Conference} on {Blockchain} ({Blockchain})}}. \bibinfo{pages}{250--257}.
\newblock
\urldef\tempurl%
\url{https://doi.org/10.1109/Blockchain55522.2022.00042}
\showDOI{\tempurl}


\bibitem[Kitchenham and Charters(2007)]%
        {kitchenham_guidelines_2007}
\bibfield{author}{\bibinfo{person}{Barbara~Ann Kitchenham} {and} \bibinfo{person}{Stuart Charters}.} \bibinfo{year}{2007}\natexlab{}.
\newblock \bibinfo{booktitle}{\emph{Guidelines for performing {Systematic} {Literature} {Reviews} in {Software} {Engineering}}}.
\newblock \bibinfo{type}{{T}echnical {R}eport} EBSE 2007-001. \bibinfo{institution}{Keele University}.
\newblock
\urldef\tempurl%
\url{https://www.elsevier.com/__data/promis_misc/525444systematicreviewsguide.pdf}
\showURL{%
\tempurl}
\newblock
\shownote{Backup Publisher: Keele University and Durham University Joint Report}.


\bibitem[Klinkmüller et~al\mbox{.}(2019)]%
        {klinkmuller_mining_2019}
\bibfield{author}{\bibinfo{person}{Christopher Klinkmüller}, \bibinfo{person}{Alexander Ponomarev}, \bibinfo{person}{An~Binh Tran}, \bibinfo{person}{Ingo Weber}, {and} \bibinfo{person}{Wil van~der Aalst}.} \bibinfo{year}{2019}\natexlab{}.
\newblock \showarticletitle{Mining {Blockchain} {Processes}: {Extracting} {Process} {Mining} {Data} from {Blockchain} {Applications}}. In \bibinfo{booktitle}{\emph{Business {Process} {Management}: {Blockchain} and {Central} and {Eastern} {Europe} {Forum}}}, \bibfield{editor}{\bibinfo{person}{Claudio Di~Ciccio}, \bibinfo{person}{Renata Gabryelczyk}, \bibinfo{person}{Luciano García-Bañuelos}, \bibinfo{person}{Tomislav Hernaus}, \bibinfo{person}{Rick Hull}, \bibinfo{person}{Mojca Indihar~Štemberger}, \bibinfo{person}{Andrea K{\textbackslash}textbackslashHo}, {and} \bibinfo{person}{Mark Staples}} (Eds.). \bibinfo{publisher}{Springer International Publishing}, \bibinfo{address}{Cham}, \bibinfo{pages}{71--86}.
\newblock
\showISBNx{978-3-030-30429-4}


\bibitem[Koshy et~al\mbox{.}(2014)]%
        {koshy_analysis_2014}
\bibfield{author}{\bibinfo{person}{Philip Koshy}, \bibinfo{person}{Diana Koshy}, {and} \bibinfo{person}{Patrick McDaniel}.} \bibinfo{year}{2014}\natexlab{}.
\newblock \showarticletitle{An {Analysis} of {Anonymity} in {Bitcoin} {Using} {P2P} {Network} {Traffic}}. In \bibinfo{booktitle}{\emph{Financial {Cryptography} and {Data} {Security}}}, \bibfield{editor}{\bibinfo{person}{Nicolas Christin} {and} \bibinfo{person}{Reihaneh Safavi-Naini}} (Eds.). \bibinfo{publisher}{Springer Berlin Heidelberg}, \bibinfo{address}{Berlin, Heidelberg}, \bibinfo{pages}{469--485}.
\newblock
\showISBNx{978-3-662-45472-5}


\bibitem[Laudon et~al\mbox{.}(2014)]%
        {laudon_management_2014}
\bibfield{author}{\bibinfo{person}{Kenneth Laudon}, \bibinfo{person}{Kenneth~C. Laudon}, {and} \bibinfo{person}{Jane~P. Laudon}.} \bibinfo{year}{2014}\natexlab{}.
\newblock \bibinfo{booktitle}{\emph{Management {Information} {Systems}: {Managing} the {Digital} {Firm}, {Global} {Edition}} (\bibinfo{edition}{13} ed.)}.
\newblock \bibinfo{publisher}{Pearson}.
\newblock
\urldef\tempurl%
\url{https://www.pearson.ch/management-information-systems-managing-the-digital-firm-global-edition}
\showURL{%
\tempurl}


\bibitem[Lee et~al\mbox{.}(2020b)]%
        {lee_machine_2020}
\bibfield{author}{\bibinfo{person}{Chaehyeon Lee}, \bibinfo{person}{Sajan Maharjan}, \bibinfo{person}{Kyungchan Ko}, \bibinfo{person}{Jongsoo Woo}, {and} \bibinfo{person}{James Won-Ki Hong}.} \bibinfo{year}{2020}\natexlab{b}.
\newblock \showarticletitle{Machine {Learning} {Based} {Bitcoin} {Address} {Classification}}. In \bibinfo{booktitle}{\emph{Blockchain and {Trustworthy} {Systems}}}, \bibfield{editor}{\bibinfo{person}{Zibin Zheng}, \bibinfo{person}{Hong-Ning Dai}, \bibinfo{person}{Xiaodong Fu}, {and} \bibinfo{person}{Benhui Chen}} (Eds.). \bibinfo{publisher}{Springer Singapore}, \bibinfo{address}{Singapore}, \bibinfo{pages}{517--531}.
\newblock
\showISBNx{978-981-15-9213-3}


\bibitem[Lee et~al\mbox{.}(2020a)]%
        {lee_measurements_2020}
\bibfield{author}{\bibinfo{person}{Xi~Tong Lee}, \bibinfo{person}{Arijit Khan}, \bibinfo{person}{Sourav Sen~Gupta}, \bibinfo{person}{Yu~Hann Ong}, {and} \bibinfo{person}{Xuan Liu}.} \bibinfo{year}{2020}\natexlab{a}.
\newblock \showarticletitle{Measurements, {Analyses}, and {Insights} on the {Entire} {Ethereum} {Blockchain} {Network}}. In \bibinfo{booktitle}{\emph{Proceedings of {The} {Web} {Conference} 2020}} \emph{(\bibinfo{series}{{WWW} '20})}. \bibinfo{publisher}{Association for Computing Machinery}, \bibinfo{address}{New York, NY, USA}, \bibinfo{pages}{155--166}.
\newblock
\showISBNx{978-1-4503-7023-3}
\urldef\tempurl%
\url{https://doi.org/10.1145/3366423.3380103}
\showDOI{\tempurl}


\bibitem[Li(2022)]%
        {li_survey_2022}
\bibfield{author}{\bibinfo{person}{Meng Li}.} \bibinfo{year}{2022}\natexlab{}.
\newblock \showarticletitle{A {Survey} on {Ethereum} {Illicit} {Detection}}. In \bibinfo{booktitle}{\emph{Artificial {Intelligence} and {Security}}}, \bibfield{editor}{\bibinfo{person}{Xingming Sun}, \bibinfo{person}{Xiaorui Zhang}, \bibinfo{person}{Zhihua Xia}, {and} \bibinfo{person}{Elisa Bertino}} (Eds.). \bibinfo{publisher}{Springer International Publishing}, \bibinfo{address}{Cham}, \bibinfo{pages}{222--232}.
\newblock
\showISBNx{978-3-031-06791-4}


\bibitem[Li and Whinston(2020)]%
        {li_analyzing_2020}
\bibfield{author}{\bibinfo{person}{Xiaofan Li} {and} \bibinfo{person}{Andrew~B. Whinston}.} \bibinfo{year}{2020}\natexlab{}.
\newblock \showarticletitle{Analyzing {Cryptocurrencies}}.
\newblock \bibinfo{journal}{\emph{Information Systems Frontiers}} \bibinfo{volume}{22}, \bibinfo{number}{1} (\bibinfo{date}{Feb.} \bibinfo{year}{2020}), \bibinfo{pages}{17--22}.
\newblock
\showISSN{1572-9419}
\urldef\tempurl%
\url{https://doi.org/10.1007/s10796-019-09966-2}
\showDOI{\tempurl}


\bibitem[Li et~al\mbox{.}(2017)]%
        {li_etherql_2017}
\bibfield{author}{\bibinfo{person}{Yang Li}, \bibinfo{person}{Kai Zheng}, \bibinfo{person}{Ying Yan}, \bibinfo{person}{Qi Liu}, {and} \bibinfo{person}{Xiaofang Zhou}.} \bibinfo{year}{2017}\natexlab{}.
\newblock \showarticletitle{{EtherQL}: {A} {Query} {Layer} for {Blockchain} {System}}. In \bibinfo{booktitle}{\emph{Database {Systems} for {Advanced} {Applications}}}, \bibfield{editor}{\bibinfo{person}{Selçuk Candan}, \bibinfo{person}{Lei Chen}, \bibinfo{person}{Torben~Bach Pedersen}, \bibinfo{person}{Lijun Chang}, {and} \bibinfo{person}{Wen Hua}} (Eds.). \bibinfo{publisher}{Springer International Publishing}, \bibinfo{address}{Cham}, \bibinfo{pages}{556--567}.
\newblock
\showISBNx{978-3-319-55699-4}


\bibitem[Lin et~al\mbox{.}(2022)]%
        {lin_systematic_2022}
\bibfield{author}{\bibinfo{person}{Chang-Yi Lin}, \bibinfo{person}{Hsiang-Kai Liao}, {and} \bibinfo{person}{Fu-Ching Tsai}.} \bibinfo{year}{2022}\natexlab{}.
\newblock \showarticletitle{A {Systematic} {Review} of {Detecting} {Illicit} {Bitcoin} {Transactions}}.
\newblock \bibinfo{journal}{\emph{Procedia Computer Science}}  \bibinfo{volume}{207} (\bibinfo{year}{2022}), \bibinfo{pages}{3217--3225}.
\newblock
\showISSN{1877-0509}
\urldef\tempurl%
\url{https://doi.org/10.1016/j.procs.2022.09.379}
\showDOI{\tempurl}


\bibitem[Liu et~al\mbox{.}(2021)]%
        {liu_knowledge_2021}
\bibfield{author}{\bibinfo{person}{Xiao~Fan Liu}, \bibinfo{person}{Xin-Jian Jiang}, \bibinfo{person}{Si-Hao Liu}, {and} \bibinfo{person}{Chi~Kong Tse}.} \bibinfo{year}{2021}\natexlab{}.
\newblock \showarticletitle{Knowledge {Discovery} in {Cryptocurrency} {Transactions}: {A} {Survey}}.
\newblock \bibinfo{journal}{\emph{IEEE Access}}  \bibinfo{volume}{9} (\bibinfo{year}{2021}), \bibinfo{pages}{37229--37254}.
\newblock
\urldef\tempurl%
\url{https://doi.org/10.1109/ACCESS.2021.3062652}
\showDOI{\tempurl}


\bibitem[Liu et~al\mbox{.}(2018)]%
        {liu_challenges_2018}
\bibfield{author}{\bibinfo{person}{Ying Liu}, \bibinfo{person}{Henry Han}, {and} \bibinfo{person}{Joan DeBello}.} \bibinfo{year}{2018}\natexlab{}.
\newblock \showarticletitle{The {Challenges} of {Business} {Analytics}: {Successes} and {Failures}}. \bibinfo{pages}{1--10}.
\newblock
\urldef\tempurl%
\url{http://hdl.handle.net/10125/49992}
\showURL{%
\tempurl}


\bibitem[Marsalek et~al\mbox{.}(2019)]%
        {marsalek_tackling_2019}
\bibfield{author}{\bibinfo{person}{Alexander Marsalek}, \bibinfo{person}{Thomas Zefferer}, \bibinfo{person}{Edona Fasllija}, {and} \bibinfo{person}{Dominik Ziegler}.} \bibinfo{year}{2019}\natexlab{}.
\newblock \showarticletitle{Tackling {Data} {Inefficiency}: {Compressing} the {Bitcoin} {Blockchain}}. In \bibinfo{booktitle}{\emph{2019 18th {IEEE} {International} {Conference} {On} {Trust}, {Security} {And} {Privacy} {In} {Computing} {And} {Communications}/13th {IEEE} {International} {Conference} {On} {Big} {Data} {Science} {And} {Engineering} ({TrustCom}/{BigDataSE})}}. \bibinfo{pages}{626--633}.
\newblock
\urldef\tempurl%
\url{https://doi.org/10.1109/TrustCom/BigDataSE.2019.00089}
\showDOI{\tempurl}


\bibitem[Mas’ud et~al\mbox{.}(2021)]%
        {masud_review_2021}
\bibfield{author}{\bibinfo{person}{Mohd~Zaki Mas’ud}, \bibinfo{person}{Aslinda Hassan}, \bibinfo{person}{Wahidah~Md. Shah}, \bibinfo{person}{Shekh~Faisal Abdul-Latip}, \bibinfo{person}{Rabiah Ahmad}, \bibinfo{person}{Aswami Ariffin}, {and} \bibinfo{person}{Zahri Yunos}.} \bibinfo{year}{2021}\natexlab{}.
\newblock \showarticletitle{A {Review} of {Digital} {Forensics} {Framework} for {Blockchain} in {Cryptocurrency} {Technology}}. In \bibinfo{booktitle}{\emph{2021 3rd {International} {Cyber} {Resilience} {Conference} ({CRC})}}. \bibinfo{pages}{1--6}.
\newblock
\urldef\tempurl%
\url{https://doi.org/10.1109/CRC50527.2021.9392563}
\showDOI{\tempurl}


\bibitem[Matzutt et~al\mbox{.}(2018)]%
        {matzutt_quantitative_2018}
\bibfield{author}{\bibinfo{person}{Roman Matzutt}, \bibinfo{person}{Jens Hiller}, \bibinfo{person}{Martin Henze}, \bibinfo{person}{Jan~Henrik Ziegeldorf}, \bibinfo{person}{Dirk Müllmann}, \bibinfo{person}{Oliver Hohlfeld}, {and} \bibinfo{person}{Klaus Wehrle}.} \bibinfo{year}{2018}\natexlab{}.
\newblock \showarticletitle{A {Quantitative} {Analysis} of the {Impact} of {Arbitrary} {Blockchain} {Content} on {Bitcoin}}. In \bibinfo{booktitle}{\emph{Financial {Cryptography} and {Data} {Security}}}, \bibfield{editor}{\bibinfo{person}{Sarah Meiklejohn} {and} \bibinfo{person}{Kazue Sako}} (Eds.). \bibinfo{publisher}{Springer Berlin Heidelberg}, \bibinfo{address}{Berlin, Heidelberg}, \bibinfo{pages}{420--438}.
\newblock
\showISBNx{978-3-662-58387-6}


\bibitem[McGinn et~al\mbox{.}(2016)]%
        {mcginn_visualizing_2016}
\bibfield{author}{\bibinfo{person}{Dan McGinn}, \bibinfo{person}{David Birch}, \bibinfo{person}{David Akroyd}, \bibinfo{person}{Miguel Molina-Solana}, \bibinfo{person}{Yike Guo}, {and} \bibinfo{person}{William~J. Knottenbelt}.} \bibinfo{year}{2016}\natexlab{}.
\newblock \showarticletitle{Visualizing {Dynamic} {Bitcoin} {Transaction} {Patterns}}.
\newblock \bibinfo{journal}{\emph{Big Data}} \bibinfo{volume}{4}, \bibinfo{number}{2} (\bibinfo{year}{2016}), \bibinfo{pages}{109--119}.
\newblock
\urldef\tempurl%
\url{https://doi.org/10.1089/big.2015.0056}
\showDOI{\tempurl}


\bibitem[McGinn et~al\mbox{.}(2018)]%
        {mcginn_towards_2018}
\bibfield{author}{\bibinfo{person}{D. McGinn}, \bibinfo{person}{D. McIlwraith}, {and} \bibinfo{person}{Y. Guo}.} \bibinfo{year}{2018}\natexlab{}.
\newblock \showarticletitle{Towards open data blockchain analytics: a {Bitcoin} perspective}.
\newblock \bibinfo{journal}{\emph{Royal Society Open Science}} \bibinfo{volume}{5}, \bibinfo{number}{8} (\bibinfo{year}{2018}), \bibinfo{pages}{180298}.
\newblock
\urldef\tempurl%
\url{https://doi.org/10.1098/rsos.180298}
\showDOI{\tempurl}


\bibitem[Miles et~al\mbox{.}(2014)]%
        {miles_qualitative_2014}
\bibfield{author}{\bibinfo{person}{M.B. Miles}, \bibinfo{person}{A.M. Huberman}, {and} \bibinfo{person}{J. Saldana}.} \bibinfo{year}{2014}\natexlab{}.
\newblock \bibinfo{booktitle}{\emph{Qualitative {Data} {Analysis}}}.
\newblock \bibinfo{publisher}{SAGE Publications}.
\newblock
\showISBNx{978-1-4522-5787-7}
\showLCCN{2013002036}
\urldef\tempurl%
\url{https://books.google.ch/books?id=3CNrUbTu6CsC}
\showURL{%
\tempurl}


\bibitem[Mizerka et~al\mbox{.}(2020)]%
        {mizerka_role_2020}
\bibfield{author}{\bibinfo{person}{Jacek Mizerka}, \bibinfo{person}{Agnieszka Stróżyńska-Szajek}, {and} \bibinfo{person}{Piotr Mizerka}.} \bibinfo{year}{2020}\natexlab{}.
\newblock \showarticletitle{The role of {Bitcoin} on developed and emerging markets – on the basis of a {Bitcoin} users graph analysis}.
\newblock \bibinfo{journal}{\emph{Finance Research Letters}}  \bibinfo{volume}{35} (\bibinfo{year}{2020}), \bibinfo{pages}{101489}.
\newblock
\showISSN{1544-6123}
\urldef\tempurl%
\url{https://doi.org/10.1016/j.frl.2020.101489}
\showDOI{\tempurl}


\bibitem[Moreira et~al\mbox{.}(2019)]%
        {moreira_general_2019}
\bibfield{author}{\bibinfo{person}{João Moreira}, \bibinfo{person}{Andre Carvalho}, {and} \bibinfo{person}{Tomás Horvath}.} \bibinfo{year}{2019}\natexlab{}.
\newblock \bibinfo{booktitle}{\emph{A {General} {Introduction} to {Data} {Analytics}}}.
\newblock \bibinfo{publisher}{Wiley}, \bibinfo{address}{Berlin}.
\newblock
\showISBNx{978-1-119-29625-6}


\bibitem[Motamed and Bahrak(2019)]%
        {motamed_quantitative_2019}
\bibfield{author}{\bibinfo{person}{Amir~Pasha Motamed} {and} \bibinfo{person}{Behnam Bahrak}.} \bibinfo{year}{2019}\natexlab{}.
\newblock \showarticletitle{Quantitative analysis of cryptocurrencies transaction graph}.
\newblock \bibinfo{journal}{\emph{Applied Network Science}} \bibinfo{volume}{4}, \bibinfo{number}{1} (\bibinfo{date}{Dec.} \bibinfo{year}{2019}), \bibinfo{pages}{131}.
\newblock
\showISSN{2364-8228}
\urldef\tempurl%
\url{https://doi.org/10.1007/s41109-019-0249-6}
\showDOI{\tempurl}


\bibitem[Möser and Narayanan(2022)]%
        {moser_resurrecting_2022}
\bibfield{author}{\bibinfo{person}{Malte Möser} {and} \bibinfo{person}{Arvind Narayanan}.} \bibinfo{year}{2022}\natexlab{}.
\newblock \showarticletitle{Resurrecting {Address} {Clustering} in {Bitcoin}}. In \bibinfo{booktitle}{\emph{Financial {Cryptography} and {Data} {Security}}}, \bibfield{editor}{\bibinfo{person}{Ittay Eyal} {and} \bibinfo{person}{Juan Garay}} (Eds.). \bibinfo{publisher}{Springer International Publishing}, \bibinfo{address}{Cham}, \bibinfo{pages}{386--403}.
\newblock
\showISBNx{978-3-031-18283-9}


\bibitem[Mühlberger et~al\mbox{.}(2019)]%
        {muhlberger_extracting_2019}
\bibfield{author}{\bibinfo{person}{Roman Mühlberger}, \bibinfo{person}{Stefan Bachhofner}, \bibinfo{person}{Claudio Di~Ciccio}, \bibinfo{person}{Luciano García-Bañuelos}, {and} \bibinfo{person}{Orlenys López-Pintado}.} \bibinfo{year}{2019}\natexlab{}.
\newblock \showarticletitle{Extracting {Event} {Logs} for {Process} {Mining} from {Data} {Stored} on the {Blockchain}}. In \bibinfo{booktitle}{\emph{Business {Process} {Management} {Workshops}}}, \bibfield{editor}{\bibinfo{person}{Chiara Di~Francescomarino}, \bibinfo{person}{Remco Dijkman}, {and} \bibinfo{person}{Uwe Zdun}} (Eds.). \bibinfo{publisher}{Springer International Publishing}, \bibinfo{address}{Cham}, \bibinfo{pages}{690--703}.
\newblock
\showISBNx{978-3-030-37453-2}


\bibitem[Nakamoto(2008)]%
        {nakamoto_bitcoin_2008}
\bibfield{author}{\bibinfo{person}{Satohsi Nakamoto}.} \bibinfo{year}{2008}\natexlab{}.
\newblock \bibinfo{title}{Bitcoin: {A} {Peer}-to-{Peer} {Electronic} {Cash} {System}}.
\newblock
\newblock
\urldef\tempurl%
\url{https://web.archive.org/web/20140320135003/https://bitcoin.org/bitcoin.pdf}
\showURL{%
\tempurl}


\bibitem[Niu et~al\mbox{.}(2021)]%
        {niu_incentive_2021}
\bibfield{author}{\bibinfo{person}{Jianyu Niu}, \bibinfo{person}{Ziyu Wang}, \bibinfo{person}{Fangyu Gai}, {and} \bibinfo{person}{Chen Feng}.} \bibinfo{year}{2021}\natexlab{}.
\newblock \showarticletitle{Incentive {Analysis} of {Bitcoin}-{NG}, {Revisited}}.
\newblock \bibinfo{journal}{\emph{SIGMETRICS Perform. Eval. Rev.}} \bibinfo{volume}{48}, \bibinfo{number}{3} (\bibinfo{date}{March} \bibinfo{year}{2021}), \bibinfo{pages}{59--60}.
\newblock
\showISSN{0163-5999}
\urldef\tempurl%
\url{https://doi.org/10.1145/3453953.3453966}
\showDOI{\tempurl}


\bibitem[Oggier et~al\mbox{.}(2018)]%
        {oggier_biva_2018}
\bibfield{author}{\bibinfo{person}{Frédérique Oggier}, \bibinfo{person}{Silivanxay Phetsouvanh}, {and} \bibinfo{person}{Anwitaman Datta}.} \bibinfo{year}{2018}\natexlab{}.
\newblock \showarticletitle{{BiVA}: {Bitcoin} {Network} {Visualization} \& {Analysis}}. In \bibinfo{booktitle}{\emph{2018 {IEEE} {International} {Conference} on {Data} {Mining} {Workshops} ({ICDMW})}}. \bibinfo{pages}{1469--1474}.
\newblock
\urldef\tempurl%
\url{https://doi.org/10.1109/ICDMW.2018.00210}
\showDOI{\tempurl}


\bibitem[Paik et~al\mbox{.}(2019)]%
        {paik_analysis_2019}
\bibfield{author}{\bibinfo{person}{Hye-Young Paik}, \bibinfo{person}{Xiwei Xu}, \bibinfo{person}{H.~M. N.~Dilum Bandara}, \bibinfo{person}{Sung~Une Lee}, {and} \bibinfo{person}{Sin~Kuang Lo}.} \bibinfo{year}{2019}\natexlab{}.
\newblock \showarticletitle{Analysis of {Data} {Management} in {Blockchain}-{Based} {Systems}: {From} {Architecture} to {Governance}}.
\newblock \bibinfo{journal}{\emph{IEEE Access}}  \bibinfo{volume}{7} (\bibinfo{year}{2019}), \bibinfo{pages}{186091--186107}.
\newblock
\urldef\tempurl%
\url{https://doi.org/10.1109/ACCESS.2019.2961404}
\showDOI{\tempurl}


\bibitem[Paré et~al\mbox{.}(2015)]%
        {pare_synthesizing_2015}
\bibfield{author}{\bibinfo{person}{Guy Paré}, \bibinfo{person}{Marie-Claude Trudel}, \bibinfo{person}{Mirou Jaana}, {and} \bibinfo{person}{Spyros Kitsiou}.} \bibinfo{year}{2015}\natexlab{}.
\newblock \showarticletitle{Synthesizing information systems knowledge: {A} typology of literature reviews}.
\newblock \bibinfo{journal}{\emph{Information \& Management}} \bibinfo{volume}{52}, \bibinfo{number}{2} (\bibinfo{date}{March} \bibinfo{year}{2015}), \bibinfo{pages}{183--199}.
\newblock
\showISSN{0378-7206}
\urldef\tempurl%
\url{https://doi.org/10.1016/j.im.2014.08.008}
\showDOI{\tempurl}


\bibitem[Pavithran and Thomas(2018)]%
        {pavithran_survey_2018}
\bibfield{author}{\bibinfo{person}{Deepa Pavithran} {and} \bibinfo{person}{Rajesh Thomas}.} \bibinfo{year}{2018}\natexlab{}.
\newblock \showarticletitle{A {Survey} on {Analyzing} {Bitcoin} {Transactions}}. In \bibinfo{booktitle}{\emph{2018 {Fifth} {HCT} {Information} {Technology} {Trends} ({ITT})}}. \bibinfo{pages}{227--231}.
\newblock
\urldef\tempurl%
\url{https://doi.org/10.1109/CTIT.2018.8649517}
\showDOI{\tempurl}


\bibitem[Peng et~al\mbox{.}(2019)]%
        {peng_vql_2019}
\bibfield{author}{\bibinfo{person}{Zhe Peng}, \bibinfo{person}{Haotian Wu}, \bibinfo{person}{Bin Xiao}, {and} \bibinfo{person}{Songtao Guo}.} \bibinfo{year}{2019}\natexlab{}.
\newblock \showarticletitle{{VQL}: {Providing} {Query} {Efficiency} and {Data} {Authenticity} in {Blockchain} {Systems}}. In \bibinfo{booktitle}{\emph{2019 {IEEE} 35th {International} {Conference} on {Data} {Engineering} {Workshops} ({ICDEW})}}. \bibinfo{pages}{1--6}.
\newblock
\urldef\tempurl%
\url{https://doi.org/10.1109/ICDEW.2019.00-44}
\showDOI{\tempurl}


\bibitem[Petroni et~al\mbox{.}(2018)]%
        {petroni_big_2018}
\bibfield{author}{\bibinfo{person}{Benedito Cristiano~A. Petroni}, \bibinfo{person}{Elisângela~Mônaco de Moraes}, {and} \bibinfo{person}{Rodrigo~Franco Gonçalves}.} \bibinfo{year}{2018}\natexlab{}.
\newblock \showarticletitle{Big {Data} {Analytics} for {Logistics} and {Distributions} {Using} {Blockchain}}. In \bibinfo{booktitle}{\emph{Advances in {Production} {Management} {Systems}. {Smart} {Manufacturing} for {Industry} 4.0}}, \bibfield{editor}{\bibinfo{person}{Ilkyeong Moon}, \bibinfo{person}{Gyu~M. Lee}, \bibinfo{person}{Jinwoo Park}, \bibinfo{person}{Dimitris Kiritsis}, {and} \bibinfo{person}{Gregor von Cieminski}} (Eds.). \bibinfo{publisher}{Springer International Publishing}, \bibinfo{address}{Cham}, \bibinfo{pages}{363--369}.
\newblock
\showISBNx{978-3-319-99707-0}


\bibitem[Pierro and Rocha(2019)]%
        {pierro_influence_2019}
\bibfield{author}{\bibinfo{person}{Giuseppe~Antonio Pierro} {and} \bibinfo{person}{Henrique Rocha}.} \bibinfo{year}{2019}\natexlab{}.
\newblock \showarticletitle{The {Influence} {Factors} on {Ethereum} {Transaction} {Fees}}. In \bibinfo{booktitle}{\emph{2019 {IEEE}/{ACM} 2nd {International} {Workshop} on {Emerging} {Trends} in {Software} {Engineering} for {Blockchain} ({WETSEB})}}. \bibinfo{pages}{24--31}.
\newblock
\urldef\tempurl%
\url{https://doi.org/10.1109/WETSEB.2019.00010}
\showDOI{\tempurl}


\bibitem[Schläfke et~al\mbox{.}(2013)]%
        {schlafke_framework_2013}
\bibfield{author}{\bibinfo{person}{Marten Schläfke}, \bibinfo{person}{Riccardo Silvi}, {and} \bibinfo{person}{Klaus Möller}.} \bibinfo{year}{2013}\natexlab{}.
\newblock \showarticletitle{A framework for business analytics in performance management}.
\newblock \bibinfo{journal}{\emph{International Journal of Productivity and Performance Management}} \bibinfo{volume}{62}, \bibinfo{number}{1} (\bibinfo{date}{Jan.} \bibinfo{year}{2013}), \bibinfo{pages}{110--122}.
\newblock
\showISSN{1741-0401}
\urldef\tempurl%
\url{https://doi.org/10.1108/17410401311285327}
\showDOI{\tempurl}


\bibitem[Sedkaoui(2020)]%
        {sedkaoui_how_2020}
\bibfield{author}{\bibinfo{person}{Soraya Sedkaoui}.} \bibinfo{year}{2020}\natexlab{}.
\newblock \showarticletitle{How data analytics is changing entrepreneurial opportunities?}
\newblock \bibinfo{journal}{\emph{International Journal of Innovation Science}} \bibinfo{volume}{10}, \bibinfo{number}{2} (\bibinfo{date}{Jan.} \bibinfo{year}{2020}), \bibinfo{pages}{274--294}.
\newblock
\showISSN{1757-2223}
\urldef\tempurl%
\url{https://doi.org/10.1108/IJIS-09-2017-0092}
\showDOI{\tempurl}
\newblock
\shownote{Publisher: Emerald Publishing Limited}.


\bibitem[Serena et~al\mbox{.}(2022)]%
        {serena_cryptocurrencies_2022}
\bibfield{author}{\bibinfo{person}{Luca Serena}, \bibinfo{person}{Stefano Ferretti}, {and} \bibinfo{person}{Gabriele D’Angelo}.} \bibinfo{year}{2022}\natexlab{}.
\newblock \showarticletitle{Cryptocurrencies activity as a complex network: {Analysis} of transactions graphs}.
\newblock \bibinfo{journal}{\emph{Peer-to-Peer Networking and Applications}} \bibinfo{volume}{15}, \bibinfo{number}{2} (\bibinfo{date}{March} \bibinfo{year}{2022}), \bibinfo{pages}{839--853}.
\newblock
\showISSN{1936-6450}
\urldef\tempurl%
\url{https://doi.org/10.1007/s12083-021-01220-4}
\showDOI{\tempurl}


\bibitem[Shen et~al\mbox{.}(2021)]%
        {shen_identity_2021}
\bibfield{author}{\bibinfo{person}{Jie Shen}, \bibinfo{person}{Jiajun Zhou}, \bibinfo{person}{Yunyi Xie}, \bibinfo{person}{Shanqing Yu}, {and} \bibinfo{person}{Qi Xuan}.} \bibinfo{year}{2021}\natexlab{}.
\newblock \showarticletitle{Identity {Inference} on {Blockchain} {Using} {Graph} {Neural} {Network}}. In \bibinfo{booktitle}{\emph{Blockchain and {Trustworthy} {Systems}}}, \bibfield{editor}{\bibinfo{person}{Hong-Ning Dai}, \bibinfo{person}{Xuanzhe Liu}, \bibinfo{person}{Daniel~Xiapu Luo}, \bibinfo{person}{Jiang Xiao}, {and} \bibinfo{person}{Xiangping Chen}} (Eds.). \bibinfo{publisher}{Springer Singapore}, \bibinfo{address}{Singapore}, \bibinfo{pages}{3--17}.
\newblock
\showISBNx{978-981-16-7993-3}


\bibitem[Shin et~al\mbox{.}(2021)]%
        {shin_survey_2021}
\bibfield{author}{\bibinfo{person}{Hye-Yeong Shin}, \bibinfo{person}{Meryam Essaid}, \bibinfo{person}{Sejin Park}, {and} \bibinfo{person}{Hongtaek Ju}.} \bibinfo{year}{2021}\natexlab{}.
\newblock \showarticletitle{A survey on public blockchain-based networks: structural differences and address clustering methods}. In \bibinfo{booktitle}{\emph{2021 22nd {Asia}-{Pacific} {Network} {Operations} and {Management} {Symposium} ({APNOMS})}}. \bibinfo{pages}{57--60}.
\newblock
\urldef\tempurl%
\url{https://doi.org/10.23919/APNOMS52696.2021.9562685}
\showDOI{\tempurl}


\bibitem[Sin and Wang(2017)]%
        {sin_bitcoin_2017}
\bibfield{author}{\bibinfo{person}{Edwin Sin} {and} \bibinfo{person}{Lipo Wang}.} \bibinfo{year}{2017}\natexlab{}.
\newblock \showarticletitle{Bitcoin price prediction using ensembles of neural networks}. In \bibinfo{booktitle}{\emph{2017 13th {International} {Conference} on {Natural} {Computation}, {Fuzzy} {Systems} and {Knowledge} {Discovery} ({ICNC}-{FSKD})}}. \bibinfo{pages}{666--671}.
\newblock
\urldef\tempurl%
\url{https://doi.org/10.1109/FSKD.2017.8393351}
\showDOI{\tempurl}


\bibitem[Singh and Malhotra(2023)]%
        {singh_review_2023}
\bibfield{author}{\bibinfo{person}{Devansh Singh} {and} \bibinfo{person}{Mrs~Vimmi Malhotra}.} \bibinfo{year}{2023}\natexlab{}.
\newblock \showarticletitle{A {Review} on the {Capability} and {Smart} {Contract} {Potential} of {Block} chain {Technology}}. In \bibinfo{booktitle}{\emph{2023 3rd {International} {Conference} on {Smart} {Data} {Intelligence} ({ICSMDI})}}. \bibinfo{pages}{80--87}.
\newblock
\urldef\tempurl%
\url{https://doi.org/10.1109/ICSMDI57622.2023.00022}
\showDOI{\tempurl}


\bibitem[Singhal et~al\mbox{.}(2021)]%
        {singhal_coalescence_2021}
\bibfield{author}{\bibinfo{person}{Tushar Singhal}, \bibinfo{person}{M.~S. Bhargavi}, {and} \bibinfo{person}{P. Hemavathi}.} \bibinfo{year}{2021}\natexlab{}.
\newblock \showarticletitle{Coalescence of {Artificial} {Intelligence} with {Blockchain}: {A} {Survey} on {Analytics} {Over} {Blockchain} {Data} in {Different} {Sectors}}. In \bibinfo{booktitle}{\emph{Emerging {Technologies} in {Data} {Mining} and {Information} {Security}}}, \bibfield{editor}{\bibinfo{person}{Aboul~Ella Hassanien}, \bibinfo{person}{Siddhartha Bhattacharyya}, \bibinfo{person}{Satyajit Chakrabati}, \bibinfo{person}{Abhishek Bhattacharya}, {and} \bibinfo{person}{Soumi Dutta}} (Eds.). \bibinfo{publisher}{Springer Singapore}, \bibinfo{address}{Singapore}, \bibinfo{pages}{703--711}.
\newblock
\showISBNx{978-981-15-9927-9}


\bibitem[Skyrius(2021)]%
        {skyrius_business_2021}
\bibfield{author}{\bibinfo{person}{Rimvydas Skyrius}.} \bibinfo{year}{2021}\natexlab{}.
\newblock \bibinfo{booktitle}{\emph{Business {Intelligence}: {A} {Comprehensive} {Approach} to {Information} {Needs}, {Technologies} and {Culture}}}.
\newblock \bibinfo{publisher}{Springer International Publishing}, \bibinfo{address}{Cham}.
\newblock
\showISBNx{978-3-030-67031-3 978-3-030-67032-0}
\urldef\tempurl%
\url{https://doi.org/10.1007/978-3-030-67032-0}
\showDOI{\tempurl}


\bibitem[Srivasthav et~al\mbox{.}(2021)]%
        {srivasthav_study_2021}
\bibfield{author}{\bibinfo{person}{Dinesh~P Srivasthav}, \bibinfo{person}{Lakshmi~Padmaja Maddali}, {and} \bibinfo{person}{R Vigneswaran}.} \bibinfo{year}{2021}\natexlab{}.
\newblock \showarticletitle{Study of {Blockchain} {Forensics} and {Analytics} tools}. In \bibinfo{booktitle}{\emph{2021 3rd {Conference} on {Blockchain} {Research} \& {Applications} for {Innovative} {Networks} and {Services} ({BRAINS})}}. \bibinfo{pages}{39--40}.
\newblock
\urldef\tempurl%
\url{https://doi.org/10.1109/BRAINS52497.2021.9569824}
\showDOI{\tempurl}


\bibitem[Sun et~al\mbox{.}(2019a)]%
        {sun_ethereum_2019}
\bibfield{author}{\bibinfo{person}{Hanyi Sun}, \bibinfo{person}{Na Ruan}, {and} \bibinfo{person}{Hanqing Liu}.} \bibinfo{year}{2019}\natexlab{a}.
\newblock \showarticletitle{Ethereum {Analysis} via {Node} {Clustering}}. In \bibinfo{booktitle}{\emph{Network and {System} {Security}}}, \bibfield{editor}{\bibinfo{person}{Joseph~K. Liu} {and} \bibinfo{person}{Xinyi Huang}} (Eds.). \bibinfo{publisher}{Springer International Publishing}, \bibinfo{address}{Cham}, \bibinfo{pages}{114--129}.
\newblock
\showISBNx{978-3-030-36938-5}


\bibitem[Sun et~al\mbox{.}(2019b)]%
        {sun_bitvis_2019}
\bibfield{author}{\bibinfo{person}{Yujing Sun}, \bibinfo{person}{Hao Xiong}, \bibinfo{person}{Siu~Ming Yiu}, {and} \bibinfo{person}{Kwok~Yan Lam}.} \bibinfo{year}{2019}\natexlab{b}.
\newblock \showarticletitle{{BitVis}: {An} {Interactive} {Visualization} {System} for {Bitcoin} {Accounts} {Analysis}}. In \bibinfo{booktitle}{\emph{2019 {Crypto} {Valley} {Conference} on {Blockchain} {Technology} ({CVCBT})}}. \bibinfo{pages}{21--25}.
\newblock
\urldef\tempurl%
\url{https://doi.org/10.1109/CVCBT.2019.000-3}
\showDOI{\tempurl}


\bibitem[Thelwall et~al\mbox{.}(2023)]%
        {thelwall_which_2023}
\bibfield{author}{\bibinfo{person}{Mike Thelwall}, \bibinfo{person}{Kayvan Kousha}, \bibinfo{person}{Emma Stuart}, \bibinfo{person}{Meiko Makita}, \bibinfo{person}{Mahshid Abdoli}, \bibinfo{person}{Paul Wilson}, {and} \bibinfo{person}{Jonathan Levitt}.} \bibinfo{year}{2023}\natexlab{}.
\newblock \showarticletitle{In which fields are citations indicators of research quality?}
\newblock \bibinfo{journal}{\emph{Journal of the Association for Information Science and Technology}} \bibinfo{volume}{74}, \bibinfo{number}{8} (\bibinfo{year}{2023}), \bibinfo{pages}{941--953}.
\newblock
\urldef\tempurl%
\url{https://doi.org/10.1002/asi.24767}
\showDOI{\tempurl}


\bibitem[Tovanich and Cazabet(2023)]%
        {tovanich_pattern_2023}
\bibfield{author}{\bibinfo{person}{Natkamon Tovanich} {and} \bibinfo{person}{Rémy Cazabet}.} \bibinfo{year}{2023}\natexlab{}.
\newblock \showarticletitle{Pattern {Analysis} of {Money} {Flows} in the {Bitcoin} {Blockchain}}. In \bibinfo{booktitle}{\emph{Complex {Networks} and {Their} {Applications} {XI}}}, \bibfield{editor}{\bibinfo{person}{Hocine Cherifi}, \bibinfo{person}{Rosario~Nunzio Mantegna}, \bibinfo{person}{Luis~M. Rocha}, \bibinfo{person}{Chantal Cherifi}, {and} \bibinfo{person}{Salvatore Miccichè}} (Eds.). \bibinfo{publisher}{Springer International Publishing}, \bibinfo{address}{Cham}, \bibinfo{pages}{443--455}.
\newblock
\showISBNx{978-3-031-21127-0}


\bibitem[Tovanich et~al\mbox{.}(2021a)]%
        {tovanich_visualization_2021}
\bibfield{author}{\bibinfo{person}{Natkamon Tovanich}, \bibinfo{person}{Nicolas Heulot}, \bibinfo{person}{Jean-Daniel Fekete}, {and} \bibinfo{person}{Petra Isenberg}.} \bibinfo{year}{2021}\natexlab{a}.
\newblock \showarticletitle{Visualization of {Blockchain} {Data}: {A} {Systematic} {Review}}.
\newblock \bibinfo{journal}{\emph{IEEE Transactions on Visualization and Computer Graphics}} \bibinfo{volume}{27}, \bibinfo{number}{7} (\bibinfo{year}{2021}), \bibinfo{pages}{3135--3152}.
\newblock
\urldef\tempurl%
\url{https://doi.org/10.1109/TVCG.2019.2963018}
\showDOI{\tempurl}


\bibitem[Tovanich et~al\mbox{.}(2021b)]%
        {tovanich_interactive_2021}
\bibfield{author}{\bibinfo{person}{Natkamon Tovanich}, \bibinfo{person}{Nicolas Soulié}, \bibinfo{person}{Nicolas Heulot}, {and} \bibinfo{person}{Petra Isenberg}.} \bibinfo{year}{2021}\natexlab{b}.
\newblock \showarticletitle{Interactive {Demo}: {Visualization} for {Bitcoin} {Mining} {Pools} {Analysis}}. In \bibinfo{booktitle}{\emph{2021 {IEEE} {International} {Conference} on {Blockchain} and {Cryptocurrency} ({ICBC})}}. \bibinfo{pages}{1--2}.
\newblock
\urldef\tempurl%
\url{https://doi.org/10.1109/ICBC51069.2021.9461124}
\showDOI{\tempurl}


\bibitem[Tovanich et~al\mbox{.}(2022)]%
        {tovanich_miningvis_2022}
\bibfield{author}{\bibinfo{person}{Natkamon Tovanich}, \bibinfo{person}{Nicolas Soulié}, \bibinfo{person}{Nicolas Heulot}, {and} \bibinfo{person}{Petra Isenberg}.} \bibinfo{year}{2022}\natexlab{}.
\newblock \showarticletitle{{MiningVis}: {Visual} {Analytics} of the {Bitcoin} {Mining} {Economy}}.
\newblock \bibinfo{journal}{\emph{IEEE Transactions on Visualization and Computer Graphics}} \bibinfo{volume}{28}, \bibinfo{number}{1} (\bibinfo{year}{2022}), \bibinfo{pages}{868--878}.
\newblock
\urldef\tempurl%
\url{https://doi.org/10.1109/TVCG.2021.3114821}
\showDOI{\tempurl}


\bibitem[Turner et~al\mbox{.}(2020)]%
        {turner_analysis_2020}
\bibfield{author}{\bibinfo{person}{Adam~Brian Turner}, \bibinfo{person}{Stephen McCombie}, {and} \bibinfo{person}{Allon~J. Uhlmann}.} \bibinfo{year}{2020}\natexlab{}.
\newblock \showarticletitle{Analysis {Techniques} for {Illicit} {Bitcoin} {Transactions}}.
\newblock \bibinfo{journal}{\emph{Frontiers in Computer Science}}  \bibinfo{volume}{2} (\bibinfo{year}{2020}).
\newblock
\showISSN{2624-9898}
\urldef\tempurl%
\url{https://doi.org/10.3389/fcomp.2020.600596}
\showDOI{\tempurl}


\bibitem[Victor(2020)]%
        {victor_address_2020}
\bibfield{author}{\bibinfo{person}{Friedhelm Victor}.} \bibinfo{year}{2020}\natexlab{}.
\newblock \showarticletitle{Address {Clustering} {Heuristics} for {Ethereum}}. In \bibinfo{booktitle}{\emph{Financial {Cryptography} and {Data} {Security}}}, \bibfield{editor}{\bibinfo{person}{Joseph Bonneau} {and} \bibinfo{person}{Nadia Heninger}} (Eds.). \bibinfo{publisher}{Springer International Publishing}, \bibinfo{address}{Cham}, \bibinfo{pages}{617--633}.
\newblock
\showISBNx{978-3-030-51280-4}


\bibitem[Vo et~al\mbox{.}(2018)]%
        {vo_research_2018}
\bibfield{author}{\bibinfo{person}{Hoang~Tam Vo}, \bibinfo{person}{Ashish Kundu}, {and} \bibinfo{person}{Mukesh Mohania}.} \bibinfo{year}{2018}\natexlab{}.
\newblock \showarticletitle{Research {Directions} in {Blockchain} {Data} {Management} and {Analytics}}. In \bibinfo{booktitle}{\emph{Proceedings of the 21st {International} {Conference} on {Extending} {Database} {Technology} ({EDBT})}}. \bibinfo{publisher}{Open Proceedings}.
\newblock
\showISBNx{978-3-89318-078-3}
\urldef\tempurl%
\url{https://doi.org/10.5441/002/edbt.2018.43}
\showDOI{\tempurl}


\bibitem[vom Brocke et~al\mbox{.}(2015)]%
        {vom_brocke_standing_2015}
\bibfield{author}{\bibinfo{person}{J. vom Brocke}, \bibinfo{person}{A. Simons}, \bibinfo{person}{K. Riemer}, \bibinfo{person}{B. Niehaves}, \bibinfo{person}{R. Plattfaut}, {and} \bibinfo{person}{A. \&~Cleven}.} \bibinfo{year}{2015}\natexlab{}.
\newblock \showarticletitle{Standing on the {Shoulders} of {Giants}: {Challenges} and {Recommendations} of {Literature} {Search} in {Information} {Systems} {Research}}.
\newblock \bibinfo{journal}{\emph{Communications of the Association for Information Systems}}  \bibinfo{volume}{37} (\bibinfo{year}{2015}), \bibinfo{pages}{205--224}.
\newblock
\urldef\tempurl%
\url{https://doi.org/10.17705/1CAIS.03709}
\showDOI{\tempurl}


\bibitem[Wang et~al\mbox{.}(2023)]%
        {wang_dissecting_2023}
\bibfield{author}{\bibinfo{person}{Canhui Wang}, \bibinfo{person}{Xiaowen Chu}, {and} \bibinfo{person}{Yang Qin}.} \bibinfo{year}{2023}\natexlab{}.
\newblock \showarticletitle{Dissecting {Mining} {Pools} of {Bitcoin} {Network}: {Measurement}, {Analysis} and {Modeling}}.
\newblock \bibinfo{journal}{\emph{IEEE Transactions on Network Science and Engineering}} \bibinfo{volume}{10}, \bibinfo{number}{1} (\bibinfo{year}{2023}), \bibinfo{pages}{398--412}.
\newblock
\urldef\tempurl%
\url{https://doi.org/10.1109/TNSE.2022.3210537}
\showDOI{\tempurl}


\bibitem[Wang et~al\mbox{.}(2019)]%
        {wang_blockchain-enabled_2019}
\bibfield{author}{\bibinfo{person}{Shuai Wang}, \bibinfo{person}{Liwei Ouyang}, \bibinfo{person}{Yong Yuan}, \bibinfo{person}{Xiaochun Ni}, \bibinfo{person}{Xuan Han}, {and} \bibinfo{person}{Fei-Yue Wang}.} \bibinfo{year}{2019}\natexlab{}.
\newblock \showarticletitle{Blockchain-{Enabled} {Smart} {Contracts}: {Architecture}, {Applications}, and {Future} {Trends}}.
\newblock \bibinfo{journal}{\emph{IEEE Transactions on Systems, Man, and Cybernetics: Systems}} \bibinfo{volume}{49}, \bibinfo{number}{11} (\bibinfo{year}{2019}), \bibinfo{pages}{2266--2277}.
\newblock
\urldef\tempurl%
\url{https://doi.org/10.1109/TSMC.2019.2895123}
\showDOI{\tempurl}


\bibitem[Webster and Watson(2002)]%
        {webster_analyzing_2002}
\bibfield{author}{\bibinfo{person}{Jane Webster} {and} \bibinfo{person}{Richard~T. Watson}.} \bibinfo{year}{2002}\natexlab{}.
\newblock \showarticletitle{Analyzing the {Past} to {Prepare} for the {Future}: {Writing} a {Literature} {Review}}.
\newblock \bibinfo{journal}{\emph{MIS Quarterly}} \bibinfo{volume}{26}, \bibinfo{number}{2} (\bibinfo{year}{2002}), \bibinfo{pages}{xiii--xxiii}.
\newblock
\showISSN{0276-7783}
\urldef\tempurl%
\url{https://www.jstor.org/stable/4132319}
\showURL{%
\tempurl}


\bibitem[Wei et~al\mbox{.}(2022)]%
        {wei_survey_2022}
\bibfield{author}{\bibinfo{person}{Qian Wei}, \bibinfo{person}{Bingzhe Li}, \bibinfo{person}{Wanli Chang}, \bibinfo{person}{Zhiping Jia}, \bibinfo{person}{Zhaoyan Shen}, {and} \bibinfo{person}{Zili Shao}.} \bibinfo{year}{2022}\natexlab{}.
\newblock \showarticletitle{A {Survey} of {Blockchain} {Data} {Management} {Systems}}.
\newblock \bibinfo{journal}{\emph{ACM Trans. Embed. Comput. Syst.}} \bibinfo{volume}{21}, \bibinfo{number}{3} (\bibinfo{date}{May} \bibinfo{year}{2022}).
\newblock
\showISSN{1539-9087}
\urldef\tempurl%
\url{https://doi.org/10.1145/3502741}
\showDOI{\tempurl}


\bibitem[Wu et~al\mbox{.}(2021)]%
        {wu_analysis_2021}
\bibfield{author}{\bibinfo{person}{Jiajing Wu}, \bibinfo{person}{Jieli Liu}, \bibinfo{person}{Yijing Zhao}, {and} \bibinfo{person}{Zibin Zheng}.} \bibinfo{year}{2021}\natexlab{}.
\newblock \showarticletitle{Analysis of cryptocurrency transactions from a network perspective: {An} overview}.
\newblock \bibinfo{journal}{\emph{Journal of Network and Computer Applications}}  \bibinfo{volume}{190} (\bibinfo{year}{2021}), \bibinfo{pages}{103139}.
\newblock
\showISSN{1084-8045}
\urldef\tempurl%
\url{https://doi.org/10.1016/j.jnca.2021.103139}
\showDOI{\tempurl}


\bibitem[Xi et~al\mbox{.}(2020)]%
        {xi_review_2020}
\bibfield{author}{\bibinfo{person}{He Xi}, \bibinfo{person}{Zhang Fan}, \bibinfo{person}{Lin Shenwen}, \bibinfo{person}{Mao Hongliang}, {and} \bibinfo{person}{He Ketai}.} \bibinfo{year}{2020}\natexlab{}.
\newblock \showarticletitle{A {Review} on {Data} {Analysis} of {Bitcoin} {Transaction} {Entity}}. In \bibinfo{booktitle}{\emph{2020 15th {IEEE} {Conference} on {Industrial} {Electronics} and {Applications} ({ICIEA})}}. \bibinfo{pages}{159--164}.
\newblock
\urldef\tempurl%
\url{https://doi.org/10.1109/ICIEA48937.2020.9248197}
\showDOI{\tempurl}


\bibitem[Yue et~al\mbox{.}(2019)]%
        {yue_bitextract_2019}
\bibfield{author}{\bibinfo{person}{Xuanwu Yue}, \bibinfo{person}{Xinhuan Shu}, \bibinfo{person}{Xinyu Zhu}, \bibinfo{person}{Xinnan Du}, \bibinfo{person}{Zheqing Yu}, \bibinfo{person}{Dimitrios Papadopoulos}, {and} \bibinfo{person}{Siyuan Liu}.} \bibinfo{year}{2019}\natexlab{}.
\newblock \showarticletitle{{BitExTract}: {Interactive} {Visualization} for {Extracting} {Bitcoin} {Exchange} {Intelligence}}.
\newblock \bibinfo{journal}{\emph{IEEE Transactions on Visualization and Computer Graphics}} \bibinfo{volume}{25}, \bibinfo{number}{1} (\bibinfo{year}{2019}), \bibinfo{pages}{162--171}.
\newblock
\urldef\tempurl%
\url{https://doi.org/10.1109/TVCG.2018.2864814}
\showDOI{\tempurl}


\bibitem[Zhang et~al\mbox{.}(2023)]%
        {zhang_survey_2023}
\bibfield{author}{\bibinfo{person}{Qizhi Zhang}, \bibinfo{person}{Yale He}, \bibinfo{person}{Ruilin Lai}, \bibinfo{person}{Zhihao Hou}, {and} \bibinfo{person}{Gansen Zhao}.} \bibinfo{year}{2023}\natexlab{}.
\newblock \showarticletitle{A survey on the efficiency, reliability, and security of data query in blockchain systems}.
\newblock \bibinfo{journal}{\emph{Future Generation Computer Systems}}  \bibinfo{volume}{145} (\bibinfo{year}{2023}), \bibinfo{pages}{303--320}.
\newblock
\showISSN{0167-739X}
\urldef\tempurl%
\url{https://doi.org/10.1016/j.future.2023.03.044}
\showDOI{\tempurl}


\bibitem[Zheng et~al\mbox{.}(2020)]%
        {zheng_xblock-eth_2020}
\bibfield{author}{\bibinfo{person}{Peilin Zheng}, \bibinfo{person}{Zibin Zheng}, \bibinfo{person}{Jiajing Wu}, {and} \bibinfo{person}{Hong-Ning Dai}.} \bibinfo{year}{2020}\natexlab{}.
\newblock \showarticletitle{{XBlock}-{ETH}: {Extracting} and {Exploring} {Blockchain} {Data} {From} {Ethereum}}.
\newblock \bibinfo{journal}{\emph{IEEE Open Journal of the Computer Society}}  \bibinfo{volume}{1} (\bibinfo{year}{2020}), \bibinfo{pages}{95--106}.
\newblock
\urldef\tempurl%
\url{https://doi.org/10.1109/OJCS.2020.2990458}
\showDOI{\tempurl}


\end{thebibliography}

\end{document}